\documentclass[12pt]{iopart}

\usepackage{epsfig,subfigure,color}
\usepackage{multirow}

\begin{document}

\title[]{Fermion Localization and Degenerate Resonances on Brane Array}

\author{Qun-Ying Xie${}^{1,3}$, Zhen-Hua Zhao${}^2$, Jie Yang${}^3$ and Ke Yang${}^{4}$\footnote{Corresponding author.}}

\address{${}^1$ School of Information Science and Engineering, Lanzhou University, Lanzhou 730000, China\\
${}^2$ Department of Applied Physics, Shandong University of Science and Technology, Qingdao, 266590, China\\
${}^3$ Institute of Theoretical Physics $\&$ Research Center of Gravitation, Lanzhou University, Lanzhou 730000, China\\
${}^4$ School of Physical Science and Technology, Southwest University, Chongqing 400715, China}
\ead{xieqy@lzu.edu.cn, zhaozhh78@sdust.edu.cn, yangjiev@lzu.edu.cn, keyang@swu.edu.cn}
\vspace{10pt}
%\begin{indented}
%\item[]July 2019
%\end{indented}

\begin{abstract}
In this work, we consider the multi-wall braneworld arisen from multi-scalar fields, and investigate the localization and resonances of spin-1/2 fermion on the multi-walls. We build two analytic multi-wall solutions with a polynomial superpotential and a modified sine-Gordon superpotential respectively. The massless fermion is the only bound state and localized between the two outermost sub-branes. The factors affecting the number of massive resonant fermions are analyzed. What interesting is that all the fermion resonant states are
non-degenerate for the cases of single- and two-walls, however, doubly-degenerate fermion resonant states emerge for the cases of three- and four-walls. This novel phenomenon could be potentially interesting in phenomenology.
\end{abstract}

\vspace{2pc}

\submitto{\CQG}
\maketitle
\section{Introduction}
\label{sec:intro}

Braneworld theory has received considerable attention during the last two decades since it is a natural way to solve some important issues in theoretical physics, such as the gauge hierarchy problem and cosmological constant problem \cite{Arkani-Hamed1998,Arkani-Hamed2000,Randall1999,Randall1999a}. In braneworld scenario, the Standard Model particles are confined on a four-dimensional object called brane, which is embedded into a higher dimension spacetime called bulk. The gravitons can escape into the extra dimension. It is naturally thought that the size of the extra dimensions must be tiny enough to hide the unobserved extra dimensions. However, as suggested in the famous Randall-Sundrum-2 (RS-2) model, where our four-dimension world is a 3-brane embedded into a five-dimensional anti-de Sitter bulk \cite{Randall1999a}, an effective four-dimensional Newtonian gravity can be recovered on the brane even the size of extra dimension is infinite.

The RS-2 is a kind of so-called thin braneworld model, because its brane tension is a delta-function put by hand along the extra dimension. However, it is widely accepted that there is a minimal length scale in the fundamental theory. So from a more realistic point of view, the brane should have a thickness, which is the so-called thick braneworld. One simplest way to smoothly generalize the RS-2 model into a thick braneworld is via introducing a canonical bulk scalar field minimally coupling to the gravity \cite{DeWolfe2000,Gremm2000a,Gremm2000,Csaki2000}. Moreover, in Refs.~\cite{Bogdanos2006,Liu2018,Liu2012a}, the authors constructed the thick brane model by including a bulk scalar field non-minimally coupling to the gravity. In Refs.~\cite{Koley2005,Bazeia2009a,PalKar2009,Liu2018a,Liu2018Cui}, the authors built thick brane models with a non-canonical bulk scalar field. The thick brane models was also considered in modified gravities, such as $f(R)$ gravity \cite{Afonso2007,Dzhunushaliev2010a,Liu2018FR,Bazeia2014,Liu2016Yu,Liu2016Zhong}, $f(T)$ gravity \cite{Yang2012b,Menezes2014,Yang2018,Liu2016},
$f(R,T)$ gravity \cite{Liu2017a}, Weyl integrable geometry \cite{Arias2002,Barbosa-Cendejas2005}, and Eddington inspired Born-Infeld gravity \cite{Liu2012,Fu2014}. In these models, the scalar field is usually a domain wall solution, which is a higher dimensional generalization of one-dimensional topologically non-trivial soliton called kink. The kink maps the boundaries of extra dimension into a set of scalar vacua.
In stead of including only one scalar field in the bulk, more bulk scalar fields were also considered to generate the thick brane configurations \cite{Kehagias2001,Bazeia2002,Bazeia2004a,Tahim2009,Bazeia2013a}, where, however, only one of the bulk scalars is the kink solution for generating the brane wall.

Nevertheless, it has been demonstrated that a single kink becomes unstable when it moves in a discrete lattice with a large velocity while a multi-kink solution remains stable \cite{Peyrard1984}. Therefore, the multi-kink braneworld is interesting and physically important. A double-kink solution can be obtained by a continuous deformation from a single-kink by running some parameters \cite{Yang2012b,Fu2014,Campos2002,Chumbes2011}. After deformation, the brane possesses an inner structure, i.e., one brane splits into two sub-branes, called as double-walls. In Ref.~\cite{Brito2014}, the authors presented a general procedure to build analytical multi-wall solutions by constructing the analytical multi-kink profiles from very smooth stepwise scalar field potentials. The single- and double-kink braneworld models and their corresponding localization property of various bulk matter fields on them are reviewed in the Refs.~\cite{Dzhunushaliev2010,Liu2017}.

However, multi-wall configuration is not only related with multi-kinks but can be also arisen from multi-scalar fields supporting the usual kink solutions. In Refs. \cite{SouzaDutra2008}, the authors built the double-wall braneworld with two kink scalars. The authors in Ref.~\cite{SouzaDutra2015} proposed a method to achieve the multi-wall models with an arbitrary number of scalar fields. Via adopting a special decomposition of the superpotential function and warp factor, this method is robust and significantly simplifies the field equations. Following their work, the authors constructed a double-wall braneworld and considering the localization of spin-1/2 fermions on this model in Ref.~\cite{Farokhtabar2017}.

In this work, following the method of Ref.~\cite{SouzaDutra2015}, we construct some multi-wall braneworld solutions with an arbitrary number of bulk scalar fields, which we dubbed as ``brane array", and investigate the localization and resonance of spin-1/2 fermion on the models. Especially, we take two-, three-, and four-wall models as examples to study the mass spectra of fermion resonances on them. It is interesting that some novel phenomenons emerge as the increase of number of sub-branes.

This paper is organized as follows. In Sec.~\ref{sec:model}, we construct the multi-wall solutions for two different forms of superpotential. In Sec.~\ref{sec:zero-localize}, we investigate the localization property of spin-1/2 Dirac fermion on the brane array. In Sec.~\ref{sec:massive-localize}, the resonant mass spectra of massive fermions are analyzed for the cases of two-, three-, and four-wall brane array. Brief conclusions are presented in the last section. By convention, the capital Latin indices $M,N=0,1,2,3,5$ and Greek indices $\mu,\nu=0,1,2,3$ label the five-dimensional and four-dimensional spacetime coordinates, respectively.

\section{The model}
\label{sec:model}

We are interested in Minkowski (flat) braneworlds, which are generated by $n$ scalar fields in five-dimensional spacetime and described by the following action
\begin{eqnarray}
\label{eq:action}
S\!=\!\int d^{5}x
        \sqrt{-g}\left[\frac{1}{2 \kappa_5}R-\sum_{i=1}^n\frac{1}{2}g^{MN}\partial_{M}\phi_{i}\partial_{N}\phi_{i}
        \!-\!V(\phi)\right],
\end{eqnarray}
where $R$ is the five-dimensional scalar curvature, $V(\phi)=V(\phi_{1},\phi_{2},\cdots,\phi_{n})$ is the scalar potential of the $n$ scalar fields, and $\kappa_5=8\pi G$  with $G$ the five-dimensional gravitational constant. We will set $\kappa_5=2$ for later convenience.

The line-element ansatz for a flat brane is given by
\begin{eqnarray}  \label{linee}
ds^{2}=g_{MN}dx^{M}dx^{N}=e^{2A(y)}\eta_{\mu\nu}dx^{\mu}dx^{\nu}+dy^{2},
\end{eqnarray}
where $\eta_{\mu\nu}$ is the usual four-dimensional Minkowski metric,  $e^{2A}$ is the warp factor, and $y$ denotes the coordinate of extra dimension. For a static flat brane, the warp factor $e^{2A}$ and scalar fields $\phi_i$ are functions of the extra dimensional coordinate $y$ only.

Using the line-element (\ref{linee}), one gets the following second-order nonlinear coupled differential field equations
\begin{eqnarray}
-\frac{3}{2}A^{\prime\prime}=\sum_{i=1}^{n}\phi_{i}^{\prime 2} ,
  \label{eq:eom1} \\
  12A^{\prime 2}+3{A''} =-4V,  \label{eq:eom2}\\
\phi_{i}^{\prime\prime}+4A^{\prime}\phi_{i}^{\prime}=\frac{dV}{d\phi_{i}}.  \label{eq:eom3}
\end{eqnarray}
It is not easy to analytically solve the above second-order field equations directly. However, one can reduce them to first-order field equations by introducing an auxiliary superpotential $W=W(\phi_1,\phi_2,\cdot\cdot\cdot,\phi_n)$\cite{DeWolfe2000,Gremm2000a,Bazeia2004a,Afonso2006}. Then, the corresponding first-order field equations is written as
\begin{eqnarray}
 A' &=&-\frac{2}{3}W, \label{1stOrder1} \\
 \phi_{i}' &=& \frac{\partial W}{\partial \phi_{i}},~~\label{1stOrder2}\\
 V &=& \frac{1}{2}\sum_{i=1}^{n}\Big(\frac{\partial W}{\partial \phi_{i}}\Big)^{2}
    -\frac{4}{3}W^{2}. \label{1stOrder3}
\end{eqnarray}
In order to solve these multi-scalar brane system, one can
 simply decompose the superpotential $W$ into a sum of superpotentials $W_{i}$ \cite{SouzaDutra2015,Farokhtabar2017}, i.e., $W = \sum_{i=1}^{n} W_{i}(\phi_{i})$, and
rewrite the warp factor  $A$ as $A=\sum_{i=1}^{n}  A_{i}(y)$ correspondingly, then the first order equations can be reduced to
\begin{eqnarray}
 A_i'(y)=-\frac{2}{3}W_i(\phi_i), \label{1stOrder1b}\\
 \phi_{i}' = \frac{\partial W_i}{\partial \phi_{i}}. \label{1stOrder2b}
\end{eqnarray}
Now, the original field equations (\ref{eq:eom1})-(\ref{eq:eom3}) are replaced by the first-order equations (\ref{1stOrder3})-(\ref{1stOrder2b}). By imposing an explicit form of superpotential, we can achieve an analytical brane solution easily.

\subsection{Polynomial superpotential}

As the first example, we consider that each superpotential $W_i$ has the same polynomial form like
\begin{eqnarray}  \label{caseI_Wi}
   W_i(\phi_i)=k v_i \left( \phi_i - \frac{\phi_i^3}{3 v_i^2} \right), ~~~(i=1,2,\cdots,n).
\end{eqnarray}
By substituting the above $W_i$ into Eqs. (\ref{1stOrder3}), (\ref{1stOrder1b}) and (\ref{1stOrder2b}), we get the solution of the scalar potential $V(\phi)$, the scalar fields $\phi_{i}$ and the warp factors $A_{i}$:
\begin{eqnarray}
   V(\phi) &=&
       \frac{1}{2}\sum_{i=1}^{n}\left[kv_i \left(1- \left(\frac{\phi_i}{v_i}\right)^2\right)\right]^{2} -\frac{4}{3}
        \left[\sum_{i=1}^{n}k v_i \left(\phi_i - \frac{\phi_i^3}{3v_i^2}\right) \right]^{2}, \label{SolutionI_V}\\
   \phi_{i}(y)&=&
          v_i \tanh \big(k (y-y_i)\big) , \label{SolutionI_phi} \\
   A_i(y) &=&
       \frac{1}{9} v_i^2
      \Big[ {\rm sech}^2(k(y-y_i))-{\rm sech}^2(k y_i) ] +4 \ln \Big(\frac{\cosh (k y_i)}{\cosh (k(y-y_i))}
                   \Big)
      \Big], \label{SolutionI_Ai}
\end{eqnarray}
where $y_{i}$ is a constant representing the center of each kink $\phi_{i}$, and the parameters $k$ and $v_i$ determine the topological charge of the scalar fields, i.e., when $k$ and $v_i$ have the same sign or the opposite signs, the solution is a kink or an anti-kink soliton, respectively.

%%%%%%%%%%%%%%%%%%%%%%%%%%%%%%%%%%%%%%%%%%%%%%%%%%%%%%%%%%%%%%
\begin{figure}[htbp]
\centering
\subfigure[~$\phi_i(y)$]{\label{fig_phi}
\includegraphics[width=6cm]{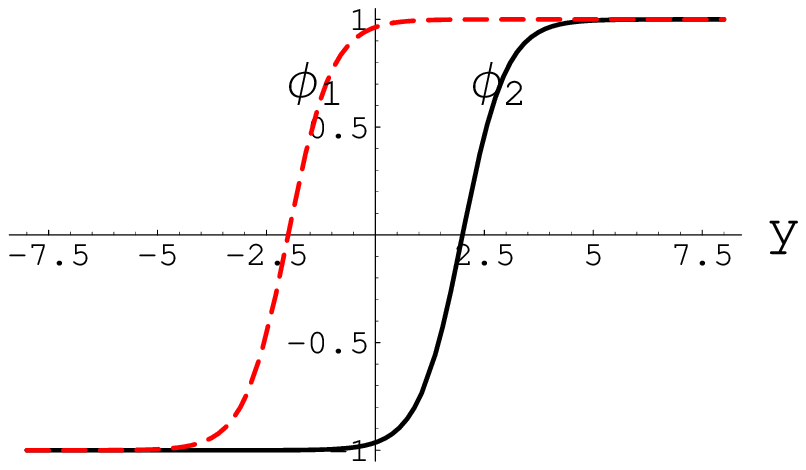}}  %from file "Plot_eAy_phi_Math5.nb"
\subfigure[~$e^{A_i(y)}$ and $e^{A(y)}$]{\label{fig_eAy}
\includegraphics[width=6cm]{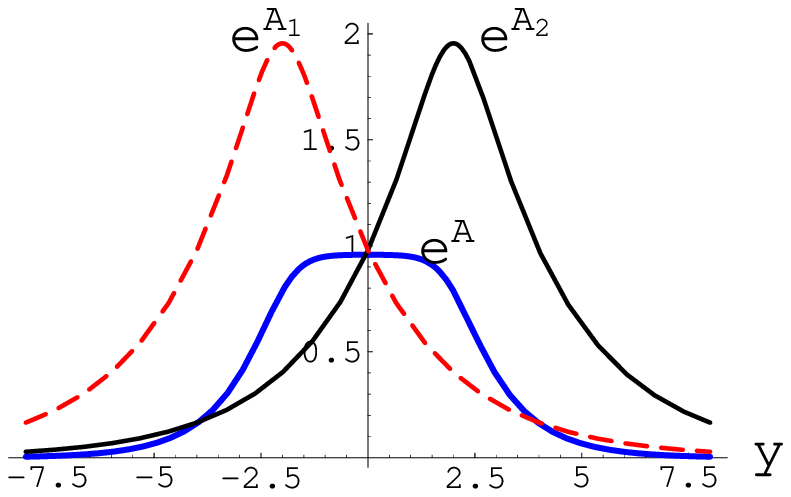}}
\caption{
\small Plots of the scalar fields $\phi_i(y)$ and the warp factors $e^{A_i(y)}$
  for $y_1=-2$ (dashed red lines),
  $y_2=2$ (thin black lines), and $k=v_1=v_2=1$. The overall warp factor
  $e^{A}$ is plotted with thick blue line.
  \label{fig_eAy_phi}}
\end{figure}
%%%%%%%%%%%%%%%%%%%%%%%%%%%%%%%%%%%%%%%%%%%%%%%%%%%%%%%%%%%%%%%

\begin{figure}[htbp]  %%figures from file rho_SolutionI_math10.nb
\centering
\subfigure[$y_1=0$]{\label{Solution1_1}
\includegraphics[width=6cm]{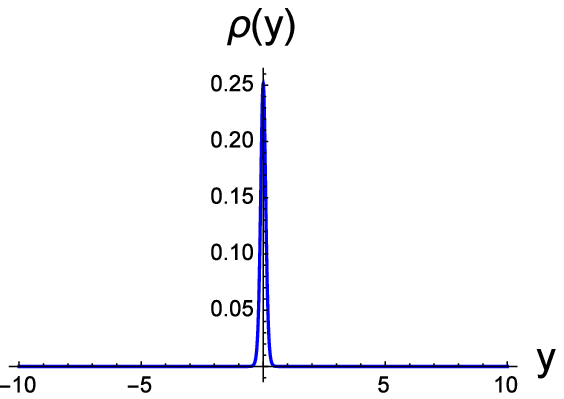}}
\subfigure[$(y_1,y_2)=(-2,2)$]{\label{Solution1_2}
\includegraphics[width=6cm]{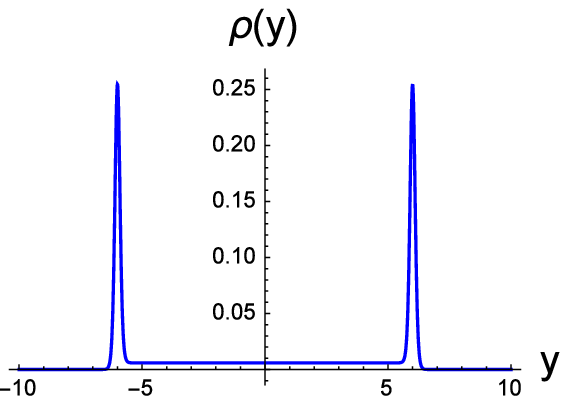}}\\
\subfigure[$(y_1,y_2,y_3)=(-2,0,2)$]{\label{Solution1_3}
\includegraphics[width=6cm]{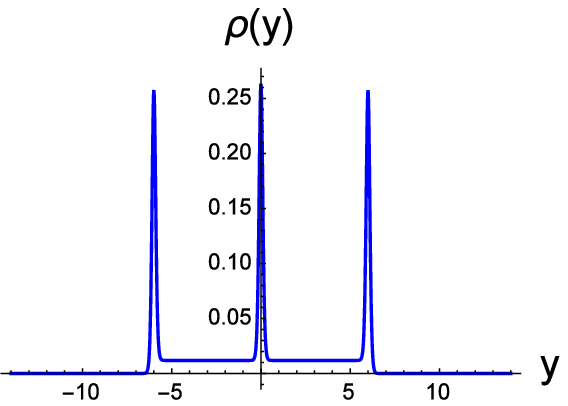}}
\subfigure[$(y_1,y_2,y_3,y_4)=(-4,-2,2,4)$]{\label{Solution1_4}
\includegraphics[width=6cm]{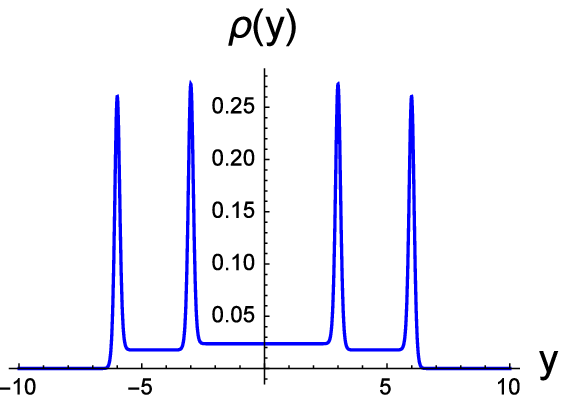}}
\caption{ \label{fig_energydensity_SolutionI}
\small Plots of the energy density $\rho(y)$ for single, double, three,  and four sub-branes. The parameters $k$ and $v_i$ are set to $k=5,v_i=0.1$ for all the sub-figures. %And $y_i$ are denoted in each sub-figure respectively.
}
\end{figure}

We find that the brane has different inner structures for different values of the parameters $k,~v_i,~y_i,$ and $n$.
The shapes of the scalar fields $\phi_i$ and the warp factors $e^{A_i(y)}$ and $e^{A(y)}$ are plotted in Fig.~\ref{fig_eAy_phi} for the case of $n=2$.
For the set of parameters with $v_1=v_2$, $y_1=-y_2$, we will obtain a thick brane consisted of two symmetric sub-branes. For the case of $n$ scalar fields with appropriate parameters, there are $n$ sub-branes. This can be easily seen from the expression of the energy density:
\begin{eqnarray}
   \rho(y)= \frac{1}{2}\sum_n \left(\partial_y \phi_i \right)^2
                 + V\left(\phi(y)\right)- V\left(\phi(y \rightarrow \infty)\right) , \label{rho_y}
\end{eqnarray}
where $V(\phi(y))$ is given by Eq. (\ref{SolutionI_V}) and $V(\phi(y \rightarrow \infty))$
is a constant decided by the five-dimensional cosmological constant.
The shapes of the energy density with one to four scalar fields are plotted in Fig.~\ref{fig_energydensity_SolutionI}.
It can be seen that, correspondingly, there are one to four sub-branes.

%%%%%%%%%%%%%%%%%%%%%%%%%%%%%%%%%%%%%%%%%%%%%%%%%%%%%%%%%%%%%%
%%%%%%%%%%%%%%%%%%%%%%%%%%%%%%%%%%%%%%%%%%%%%%%%%%%%%%%%%%%%%%%

\subsection{Modified sine-Gordon superpotential}
%$W_i(\phi_i)=\frac{k v_i}{\sqrt{3}}\Big(\phi_i+
%              v_i \sin(\frac{2\phi_i}{v_i}) \Big)$

\begin{figure}[htbp] %figures from file rho_SolutionII_math10.nb
\centering
\subfigure[$y_1=0$]{\label{Solution2_1}
\includegraphics[width=6cm]{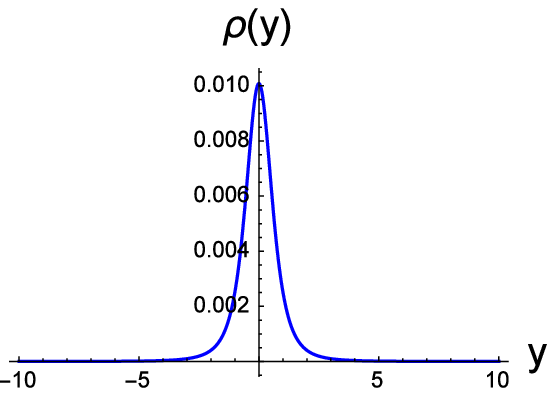}}
\subfigure[$(y_1,y_2)=(-2,2)$]{\label{Solution2_2}
\includegraphics[width=6cm]{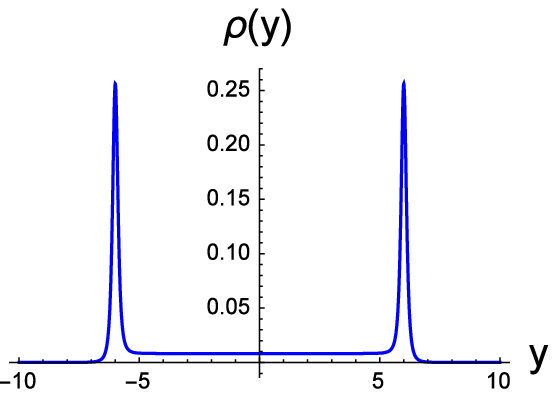}}
\subfigure[$(y_1,y_2,y_3)=(-2,0,2)$]{\label{Solution2_3}
\includegraphics[width=6cm]{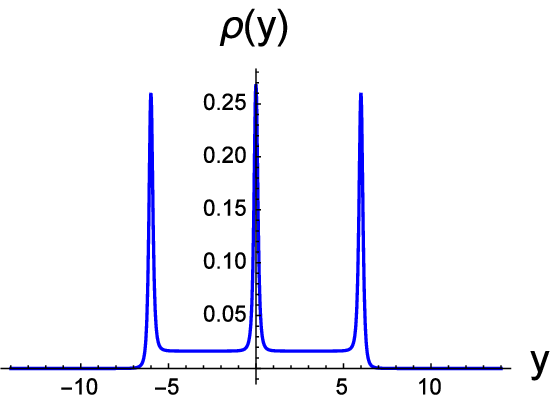}}
\subfigure[$(y_1,y_2,y_3,y_4)=(-4,-2,2,4)$]{\label{Solution2_4}
\includegraphics[width=6cm]{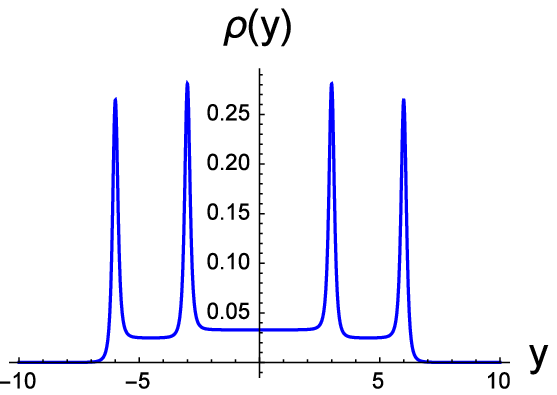}}
\caption{ \label{fig_energydensity_SolutionII}
\small Plots of the energy density $\rho(y)$ for single, double, three,
  and four sub-branes.
The parameters $k$ and $v_i$ are set to $k=5,v_i=0.05$ for all the
sub-figures. %And $y_i$ are denoted in each sub-figure respectively.
}
\end{figure}

Next, we consider another braneworld model with the following modified sine-Gordon superpotential
\begin{eqnarray} \label{eq:superpotentials_2}
  W_i(\phi_i)=k v_i \left(\phi_i+  v_i \sin \frac{\phi_i}{ v_i} \right),  ~~~(i=1,2,\cdots,n).
\end{eqnarray}
The corresponding solution is given by
\begin{eqnarray}
   V(\phi) &=&
       \frac{1}{2}\sum_{i=1}^{n}\left[k v_i \left(1+
       \cos \frac{\phi_i}{v_i}\right) \right]^{2} -\frac{4}{3}
        \left[\sum_{i=1}^{n}k v_i \left(\phi_i +
         v_i \sin \frac{\phi_i}{v_i}\right)
         \right] ^{2}, \label{SolutionII_V}\\
 {\phi_i}(y)
   &=& 2 v_i \arctan \big(k (y-y_i)\big), \label{SolutionII_phi}  \\
   {A_i}(y)
   &=& -\frac{4}{3}k v_i^2 y \arctan \big(k (y-y_i)\big).  \label{SolutionII_A}
\end{eqnarray}
The shapes of the energy density with one to four scalar fields for the modified sine-Gordon superpotential are plotted in Fig.~\ref{fig_energydensity_SolutionII}. Comparing with
Fig.~\ref{fig_energydensity_SolutionI}, we can see that, the brane configurations are quite similar. Therefore, we only focus on the first solution (\ref{caseI_Wi}) to illustrate the localization property of the fermion resonance in the following discussions.

%%%%%%%%%%%%%%%%%%%%%%%%%%%%%%%%%%%%%%%%%%%%%%%%%%%%%%%%%%%%%%
%%%%%%%%%%%%%%%%%%%%%%%%%%%%%%%%%%%%%%%%%%%%%%%%%%%%%%%%%%%%%%%

%%%%%%%%%%%%%%%%%%%%Localalization of fermionic zero mode
\section{Localization of massless fermions}
\label{sec:zero-localize}

In this section, we would like to consider the localization property of spin-1/2 Dirac fermion on the brane array. It has been proven that in order to have a normalizable fermion zero mode, the Dirac spinor field should couple with the background scalars \cite{Randjbar-Daemi2000}. Here we introduce an usual Yukawa-type interaction, and the corresponding action reads
\begin{eqnarray}
\label{eq:fermion_action}
S_{\frac{1}{2}}=\int d^{5}x \sqrt{-g} [\bar{\Psi}\Gamma^{M}D_{M}\Psi-\eta \bar{\Psi}F(\phi)\Psi],
\end{eqnarray}
where $\eta$ is the Yukawa coupling constant and $F(\phi)$  $\equiv $  $F(\phi_1,\phi_2,\cdots,\phi_n)$.

By using a coordinate transformation
\begin{eqnarray}\label{conformally_flat_coordinate}
  dz=e^{-A}dy,
\end{eqnarray}
one get a conformally flat metric
\begin{eqnarray}
\label{conformally_flat_metric}
ds^{2}=e^{2A(z)}(\eta_{\mu\nu}dx^{\mu}dx^{\nu}+dz^{2}).
\end{eqnarray}
With the conformal metric, the non-vanishing components of the spin connection is given by $\omega_\mu=\frac{1}{2}(\partial_z A)\gamma_\mu\gamma_5$. Then, the five-dimensional Dirac equation reads explicitly as
\begin{eqnarray}
 \big[ \gamma^{\mu}\partial_{\mu}
         + \gamma^5 \left(\partial_z  +2  A'(z) \right)
         -\eta e^A F(\phi)
 \big ] \Psi =0. \label{DiracEq1}
\end{eqnarray}
Considering the chirality of the Dirac field, the corresponding Kaluza-Klein (KK) decomposition is given by
\begin{eqnarray}
 \Psi(x,z) &=& \sum_n \Big(
       \psi_{{\rm L}n}(x) \hat{f}_{{\rm L}n}(z)
       \!+\!\psi_{{\rm R}n}(x) \hat{f}_{{\rm R}n}(z)
       \Big) \nonumber \\
 &=& e^{-2A}\sum_n  \Big(  \psi_{{\rm L}n}(x) f_{{\rm L}n}(z)
       +\psi_{{\rm R}n}(x) f_{{\rm R}n}(z) \Big),
\end{eqnarray}
where
\begin{eqnarray}  \label{fL_psiL}
 \hat{f}_{\rm Ln,Rn}(z)={e}^{-2A}{f}_{{\rm L},{\rm R}}(z),\quad
 \psi_{{\rm L}n,{\rm R}n}(x)=\mp\gamma^5 \psi_{{\rm L}n,{\rm R}n}(x).
\end{eqnarray}
It can be shown that the left- and right-chiral fermions satisfy the
4-dimensional Dirac equations
$\gamma^{\mu}\partial_{\mu}\psi_{{\rm L}n,{\rm R}n}=m_n\psi_{{\rm R}n,{\rm L}n}$, and the KK modes
satisfy the following coupled equations:
\begin{eqnarray}
 \left[\partial_z +\eta e^A F(\phi) \right]f_{{\rm L}n}(z)
  &\!=&\!  +m_n f_{{\rm R}n}(z), \label{CoupleEq1a}  \\
 \left[\partial_z -\eta e^A F(\phi) \right]f_{{\rm R}n}(z)
  &\!=&\!  -m_n f_{{\rm L}n}(z). \label{CoupleEq1b}
\end{eqnarray}
The above coupled equations can be rewritten into the following
Schr\"{o}dinger-like equations:
\begin{eqnarray}
  \left[-\partial_z^2 + V_{\rm L}(z) \right]f_{{\rm L}n}
            &\!=&\! m_n^2 f_{{\rm L}n},
   \label{SchEqLeftFermion}  \\
  \left[-\partial^2_z + V_{\rm R}(z) \right]f_{{\rm R}n}
            &\!=&\! m_n^2 f_{{\rm R}n},
   \label{SchEqRightFermion}
\end{eqnarray}
where the effective potentials for the left- and right-chiral KK
modes are given by
\begin{eqnarray}
  V_{{\rm L},{\rm R}}(z)= \left(\eta e^A F(\phi)\right)^2
            \mp \partial_z \big(\eta e^A F(\phi)\big).  \label{VLfermion}
\end{eqnarray}
Furthermore, by introducing an operators $\mathcal{K}$ and its conjugation $\mathcal{K}^\dag$ as $\mathcal{K}=-\partial_z+\eta e^A F(\phi)$ and $\mathcal{K}^\dag=\partial_z+\eta e^A F(\phi)$, the Hamiltonians of left- and right-chiral fermions can be rewritten into a form of supersymmetric quantum mechanics as $H_{{\rm L}}=\mathcal{K}\mathcal{K}^\dag$ and $H_{{\rm R}}=\mathcal{K}^\dag\mathcal{K}$. So all the eigenvalues, i.e., the mass squares $m_n^2$, are non-negative, and the eigenspectra of these two Hamiltonians are exactly the same except for the ground state. The wave functions of the ground states can be easily obtained by setting $m_{n}=0$ in Eqs.~(\ref{CoupleEq1a}) and (\ref{CoupleEq1b}), namely, $\mathcal{K}^\dag f_{{\rm L}n}=0$ and $\mathcal{K} f_{{\rm R}n}=0$. Then, the left- and right-chiral KK fermion zero modes read explicitly as
\begin{eqnarray}
    f_{{\rm L}0,{\rm R}0}\propto \exp \left[\mp\eta\int_{0}^{z}dz^{\prime}\mathrm{e}^{A(z^{\prime})}F(\phi) \right]. \label{fL0_z}
\end{eqnarray}
However, it implies that at most only one of the two massless modes can be localized on the brane depending on the sign of the coupling constant $\eta$. Since $V_{\rm L}(z)$ and $V_{\rm R}(z)$ are partner potentials and all the non-null eigenvalues are equal in both chiralities, we will only consider the left-chiral potential $V_{\rm L}(z)$ with the positive $\eta$ without loss of generality.

The normalization condition for the left-chiral zero mode is
\begin{eqnarray} \label{norma_condition_z}
\int_{-\infty}^{\infty}dz\exp
\left(
-2\eta \int_{0}^{z}dz^{\prime}e^{A(z^{\prime})}F \big(\phi(z')\big)
\right)
< \infty.
\end{eqnarray}
It is convenient to deal with the integral in the $y$ coordinate since
the analytic expressions of $A(y)$ and $\phi(y)$. With the coordinate transformation (\ref{conformally_flat_coordinate}),
we can rewrite the normalization condition (\ref{norma_condition_z}) to be
\begin{eqnarray}
\int_{-\infty}^{\infty}dy ~  \mathcal{I}(y)  < \infty,  \label{norma_condition_y}
\end{eqnarray}
with the integrand
\begin{eqnarray}
\mathcal{I}(y) \equiv \exp \Big(-A(y)-2\eta \int_{0}^{y}dy'
                          F \big(\phi(y')\big) \Big). \label{I}
\end{eqnarray}
From this equation, it is clear that the introduction of the scalar-fermion coupling is necessary for the localization the fermion zero mode on the brane array, since the function $e^{-A(y)}$ is divergent when $y\rightarrow\pm \infty $. In order to study the localization of the fermion zero mode on the brane array, as an example, we choose the scalar function $F(\phi)$ as a simple and natural form, namely, $F(\phi)=\phi_{1}+\phi_{2}+\cdots+\phi_{n}$. Then,
substituting the solution (\ref{SolutionI_phi}) of $\phi_{i}$ into $F(\phi)$, the integrand in (\ref{I}) can be expressed as
\begin{eqnarray} \label{I0_1}
\mathcal{I}(y)
      &=&\prod_{i=1}^{n} \exp \Bigg[
             -\frac{1}{9} v_i^2 \bigg( {\rm sech}^2(k(y-y_i))
                            -{\rm sech}^2(k y_i)\nonumber\\
      &&+4 \ln \left(\frac{\cosh (k y_i)}{\cosh (k(y-y_i))}\right)
              \bigg)  -2\eta \int_{0}^{y}dy' v_i \tanh(k(y'-y_i))\Bigg].
\end{eqnarray}
For easier illustration, all the parameters $k$ and $v_i$ are set to be  positive, i.e., we only consider kink solutions for all the scalar fields $\phi_i$. Thus, the asymptotic behavior of $\mathcal{I}$ is given by
\begin{eqnarray}
\mathcal{I}(y\rightarrow \pm\infty) \rightarrow
  ~\prod_i \exp\left[\left( \frac{4}{9}  k v_i^2 - 2\eta v_i \right)|y|\right].
\end{eqnarray}
Then, the normalization condition of the left-chiral fermion zero mode is turned out to be
\begin{eqnarray}
   \eta>\eta_{{\rm c}} \equiv \frac{2}{9} \frac{\sum_{i} k v_i^2}{ \sum_i v_i}.
\end{eqnarray}
This means that only if the Yukawa interaction is stronger enough, i.e., the coupling constant $\eta$ larger than the critical value $\eta_{{\rm c}}$, can the left-chiral fermion zero mode be ``glued" on the brane array.

By using the coordinate transformation (\ref{conformally_flat_coordinate}), we have the explicit forms of the effective potentials (\ref{VLfermion}) in the $y$ coordinate, namely,
\begin{eqnarray}
  V_{\rm L,R}(y)&=& \eta e^{2A} \Big(
            \eta F^2(\phi)- F(\phi) \partial_y A
                -\partial_y F(\phi)
            \Big),  \nonumber\\
        &=&\eta  e^{2A}
           \left(
                \eta \Big(\sum_i \phi_i \Big)^2
                \mp \Big(\sum_i \phi_i\Big) \Big(\sum_i\partial_y  A_i\Big) \mp \sum_i \partial_y \phi_i
            \right).   \label{VLfermion_explicit}
\end{eqnarray}
For example, if we consider the parameters as
\begin{eqnarray}
 v_1 &=& v_2=\cdots=v_n=v,\\
 y_1 &=& -y_n, ~~~y_2=-y_{n-1},~~~\cdots,
\end{eqnarray}
the values of the potentials for the left- and right-chiral fermions at $y=0$ can be simplified as
\begin{eqnarray}
 V_{\rm L}(0) &=&-V_{\rm R}(0) =\left\{ \begin{array}{ll}
  -2kv\eta \sum_{i=1}^{\frac{n}{2}}  {\rm sech}^2(ky_i),
         \qquad\qquad  \quad  {\rm even~ n}, \\
  -kv\eta\Big(1+ 2\sum_{i=1}^{\frac{n-1}{2}} {\rm sech}^2(ky_i)\Big),
          \qquad{\rm odd~ n}.
\end{array} \right.
\end{eqnarray}
It is clear that the values of the two potentials at $y=0$ are always opposite, and this is consistent with the conclusion that only one chiral zero mode is normalizable. The potential $V_{\rm L}(y)$ vanishes at the boundaries $y \rightarrow\pm \infty$ of the extra dimension, which is independent of the number of the
scalars $n$, the coupling constant $\eta$, and the parameters $k$ and $v$. So only the massless zero mode is bound state, and all the massive states are unbounded.

%%%%%%%%%%%%%%%%%%%%%%%%%%%%%%%%%%%%%%%%%%%%%%%%%%%%%%%%%%%%%%
\begin{figure}[htbp]
\centering
\subfigure[$v_2=0.2,~\eta=1$]{
%v1=v3=1,v2=0.2,\eta=1
\label{fig_potential_two_three_brane_a}
\includegraphics[width=5cm]{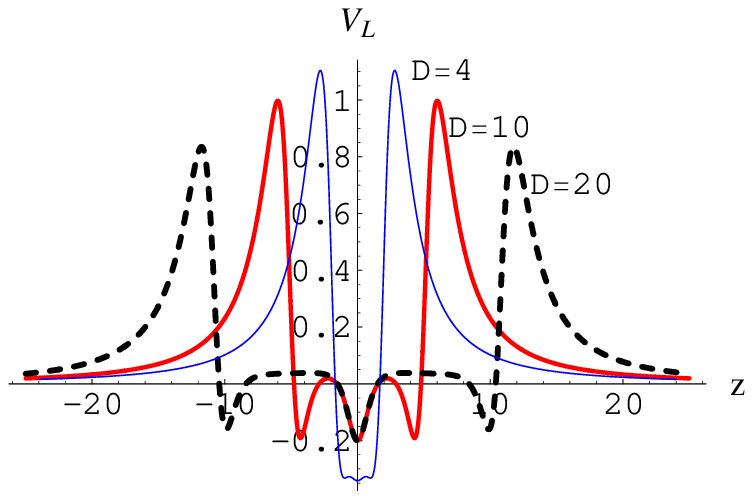}}
\subfigure[$v_2=0.2,~D=20$]
{%v1=v3=1,v2=0.2,d=20
\label{fig_potential_two_three_brane_b}
\includegraphics[width=5cm]{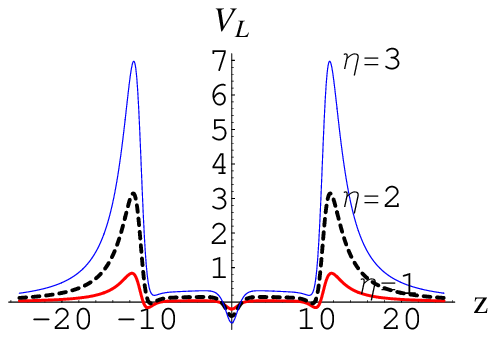}}
\subfigure[$\eta=2,~D=20$]{
%v1=v3= 1,\eta=2
\label{fig_potential_two_three_brane_c}
\includegraphics[width=5cm]{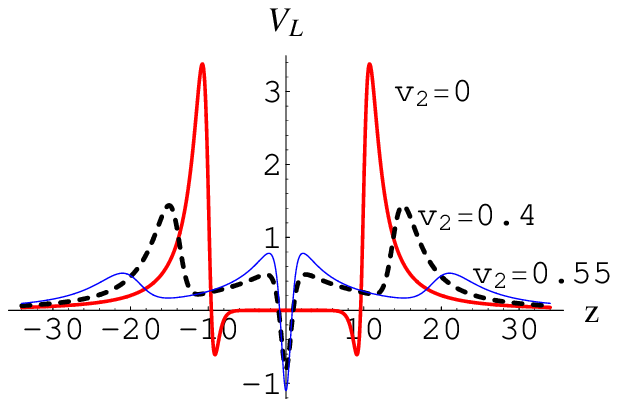}}
\caption{ \label{fig_potential_two_three_brane}
\small Plots of three-well potentials. The parameters are set to
  $k=1, v_1=v_3=1$ for all the sub-figures.
  The other parameters are set to $v_2=0.2,~\eta=1$ for (a), $v_2=0.2,~D=20$ for (b),
  and $\eta=2,~D=20$ for (c). }
\end{figure}
%%%%%%%%%%%%%%%%%%%%%%%%%%%%%%%%%%%%%%%%%%%%%%%%%%%%%%%%%%%%%%%

Here, in order to reveal how the parameters affect the potential $V_L$, we take the three-wall brane ($n=3$) as an example. It is convenient to define the total width of the brane array as
$D=|y_n-y_1|$,
and the spacing of two adjacent sub-branes as $d$.
The typical shapes of the three-wall brane are plotted in Fig.~\ref{fig_potential_two_three_brane}. There are three inner wells in the potential.
As shown in Fig.~\ref{fig_potential_two_three_brane_a}, the width of the potential well increases with the parameter $D$, but the depth of  the potential well decreases with it.
As shown in Fig.~\ref{fig_potential_two_three_brane_b}, the depth of  the potential well increases with the coupling constant $\eta$. Moreover, if we fix $v_1=v_3$ but leave $v_2$ free, then as shown in Fig.~\ref{fig_potential_two_three_brane_c}, it is interesting to see that the potential barriers will decrease with $v_2$. Especially, for $v_2=0$ the center inner potential well vanishes since the brane array reduces to a two-wall brane in this case.

\begin{figure}[htbp]
\centering
\subfigure[$n=2,~y_1,~y_2=-6,6$]{\label{fig_fL0_hfL0_a}
\includegraphics[width=5cm]{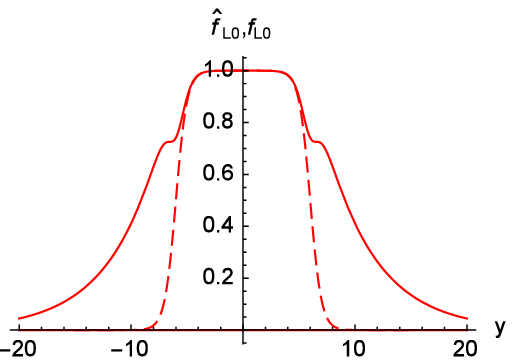}}
\subfigure[$n=3,~y_1,~y_2,~y_3=-6,0,6$]{\label{fig_fL0_hfL0_b}
\includegraphics[width=5cm]{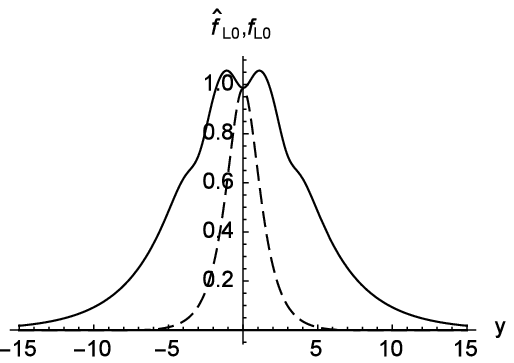}}
\subfigure[$n=4,~y_1,~y_2,~y_3,~y_4=-6,-3,3,6$]{\label{fig_fL0_hfL0_c}
\includegraphics[width=5cm]{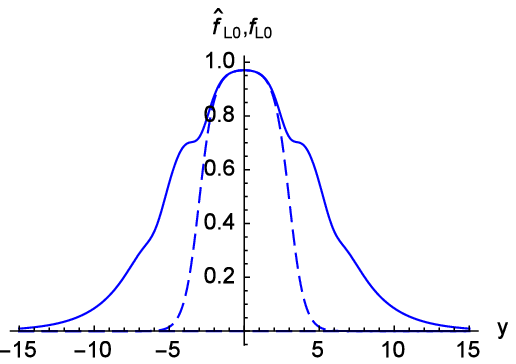}}
\caption{ \label{fig_fL0_hfL0}
\small Plots of the fermion zero modes $\hat{f}_{{\rm L}0}$ (solid line) and $f_{{\rm L}0}$ (dashed line) when there are two, three and four sub-branes. The parameters are set to $k=1, v=1, \eta=1$.}
\end{figure}

As there are many sub-branes in brane array, it is natural to specify where the fermion zero mode is localized.
From Eqs.~(\ref{conformally_flat_coordinate}) and (\ref{fL0_z}), we have
\begin{eqnarray}
  f_{{\rm L}0}
    &&\propto
         \exp\left[-\eta\int_{0}^{z}dz' e^{A(z')}F\left(\phi(z')\right)\right] \propto
         \exp \left[-\eta\int_{0}^{y}dy'F\left(\phi(y')\right)\right].
\end{eqnarray}
Figure \ref{fig_fL0_hfL0} shows $f_{{\rm L}0}$ and $\hat{f}_{{\rm L}0}={e}^{-2A}f_{{\rm L}0}$ in physical coordinate $y$ for the cases of two-, three-, and four-walls, respectively. It tells us that when there are many sub-branes, the fermion zero mode is localized between the two outermost sub-branes, whose locations are determined by the parameters $|y_1|$ and $|y_n|$. Moreover, every sub-brane will cause an inflection point in the overall shapes of fermion zero mode  $\hat{f}_{{\rm L}0}$, such as, there are two inflection points in Fig.~\ref{fig_fL0_hfL0_a}, there are three inflection points in Fig.~\ref{fig_fL0_hfL0_b}, and there are four inflection points in Fig.~\ref{fig_fL0_hfL0_c}, respectively. And the positions of the inflection points are just the centers of sub-branes.

%%%%%%%%%%%%%%%%%%%%%%%%%%%%%%%%%%%%%%%%%%%%%%%%%%%%%%%%%%%%%%
%%%%%%%%%%%%%%%%%%%%%%%%%%%%%%%%%%%%%%%%%%%%%%%%%%%%%%%%%%%%%%%

\begin{figure}[htbp]
\centering
\includegraphics[width=7cm]{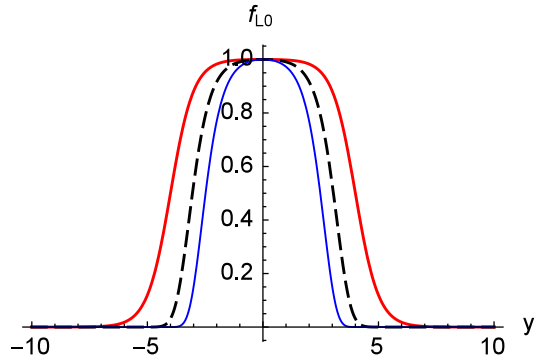}
\caption{ \label{fig_fL0}
\small Plots of fermion zero modes $f_{{\rm L}0}$ with two scalar fields.
The parameters are set to $k=1,v=1$, $\eta=1,5,15$ for red thick solid, black dashed
and  blue thin solid lines, respectively.}
\end{figure}

Furthermore, we plot the left-chiral massless fermion $f_{{\rm L}0}$ with respect to different values of coupling constant $\eta$ for the case of two sub-branes ($n=2$) in Fig.~\ref{fig_fL0}. It shows that, as expected, the wave function is more concentrated as increase of the Yukawa interaction strength.

%%%%%%%%%%%%%%%%%%%%%%%%%%%%%%%%%%%%%%%%%%%%%%%%%%%%%%%%%%%%%%
%%%%%%%%%%%%%%%%%%%%%%%%%%%%%%%%%%%%%%%%%%%%%%%%%%%%%%%%%%%%%%%

\section{Resonances of massive fermions}
\label{sec:massive-localize}

The continuous massive modes are unbounded and will propagate along the extra dimension. However, as shown in Fig.~\ref{fig_potential_two_three_brane}, since the existence of potential barriers, there could be some modes with proper eigenvalues that can stay on the brane for a long time before escaping into the extra dimension. These fermions are called quasi-localized or resonant fermions.
In Ref.~\cite{Xie2013}, the influences of some parameters on the
resonant fermions on two symmetric and asymmetric sub-branes are analyzed. Here, we focus on studying the property of resonant
mass spectra in the brane array.

Following the relative probability method presented in Ref.~\cite{Almeida2009,Liu2009c}, we calculate the probability of the continuous massive modes in the brane array.
In order to numerically solve the solutions of the KK modes $f_{{\rm L}n,{\rm R}n}$ in Eqs. (\ref{SchEqLeftFermion}) and (\ref{SchEqRightFermion}), we impose two kinds of boundary conditions, i.e.,
\begin{eqnarray}
\begin{array}{ll}
f_{{\rm L}n,{\rm R}n}(0)=1, ~{\rm and}~ f'_{{\rm L}n,{\rm R}n}(0)=0,~
{\rm for~even~parity},\\
f_{{\rm L}n,{\rm R}n}(0)=0, ~{\rm and}~ f'_{{\rm L}n,{\rm R}n}(0)=1,~
 {\rm for~odd~parity}.
  \label{Boundary_conditions}
 \end{array}
\end{eqnarray}

Since the massive modes can not be normalized, it is useful to introduce the relative probability defined as the ratio of probability in interval  $(-z_b,z_b)$ to the probability in interval $(-10z_b, 10 z_b)$, with $z_b$ the coordinate referring to the maximum of the potential \cite{Almeida2009}, namely,
\begin{eqnarray}
 P_{\rm L,R}(m)=\frac{\int_{-z_b}^{z_b} |h(z)|^2 dz}
                 {\int_{-10z_b}^{10z_b} |h(z)|^2 dz},
 \label{Probability}
\end{eqnarray}
where $|h(z)|^2$ represents the probability for finding the
massive KK mode at the position $z$ along the extra dimension.
Via scanning the relative probability for different mass, those peaks of the curve $P_{\rm L,R}(m)$ indicate the fermion resonant states. Note that the KK modes with $m^2\gg V_{\rm L,R}^{max}$ will be approximately plane waves and their solutions are $h(z)\propto\cos mz$ or $\sin mz$, so the corresponding relative probabilities $P_{\rm L,R}(m)$ trends to $0.1$.
Their lifetime can be estimated by $\tau\sim\Gamma^{-1}$, with $\Gamma=\delta m$ the width of the half height of the resonant.

\subsection{Number of resonant fermions}

Since $V_{\rm L}(z)$ and $V_{\rm R}(z)$ are partner potentials, we focus on the resonant KK modes of left-chiral fermions here, and the resonant masses of right-chiral fermions are equal to that of left-chiral fermions. Here, we investigate the effects of some factors on the relative probability $P_{\rm L}(m)$, namely, the Yukawa coupling constant $\eta$, the amplitude parameters $v_i$ of the scalar fields, the spacing $d$ between the two adjacent sub-branes, the total width $D$ of the brane array, and the number $n$ of the sub-branes.

Here, in order to investigate the effects of the amplitude parameters $v_i$ and the Yukawa coupling constant $\eta$ on the number of the resonant KK modes, as an example, we set the number of sub-branes as three, and fix the amplitude parameters of outer sub-branes as $v_1=v_3=1$ but leave $v_2$ of middle brane as a variable.

%%%%%%%%%%%%%%%%%%%%%%%%%%%%%%%%%%%%%%%%%%%%%%%%%%%%%%%%%%%%%%
\begin{figure}[htbp]
\centering
\subfigure[$D=10$]{  \label{fig_PL_v2a}
\includegraphics[width=6cm]{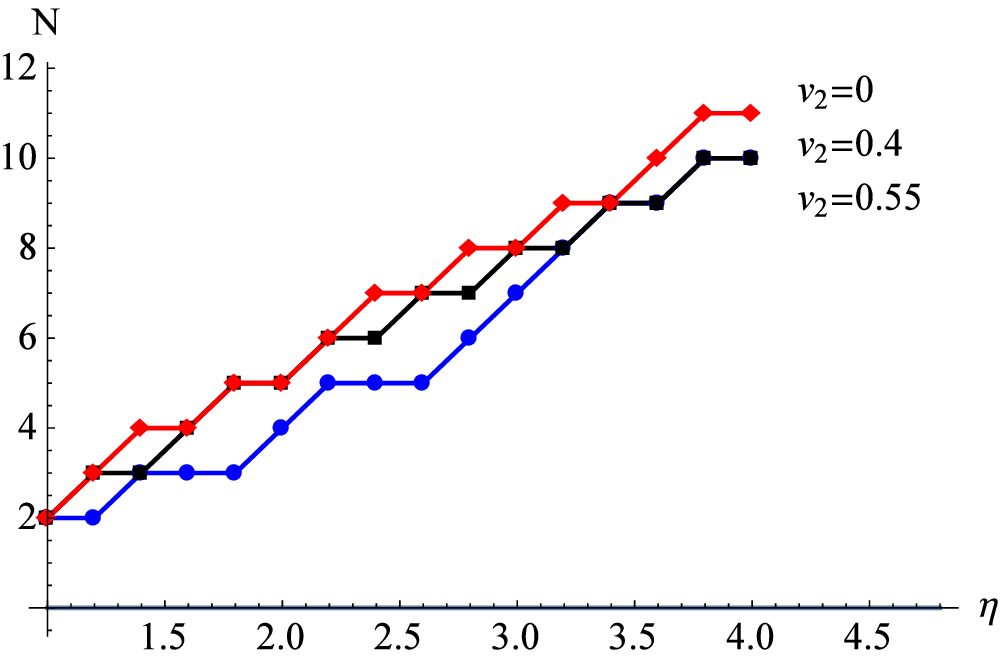}}
\subfigure[$D=20$]{   \label{fig_PL_v2b}
\includegraphics[width=6cm]{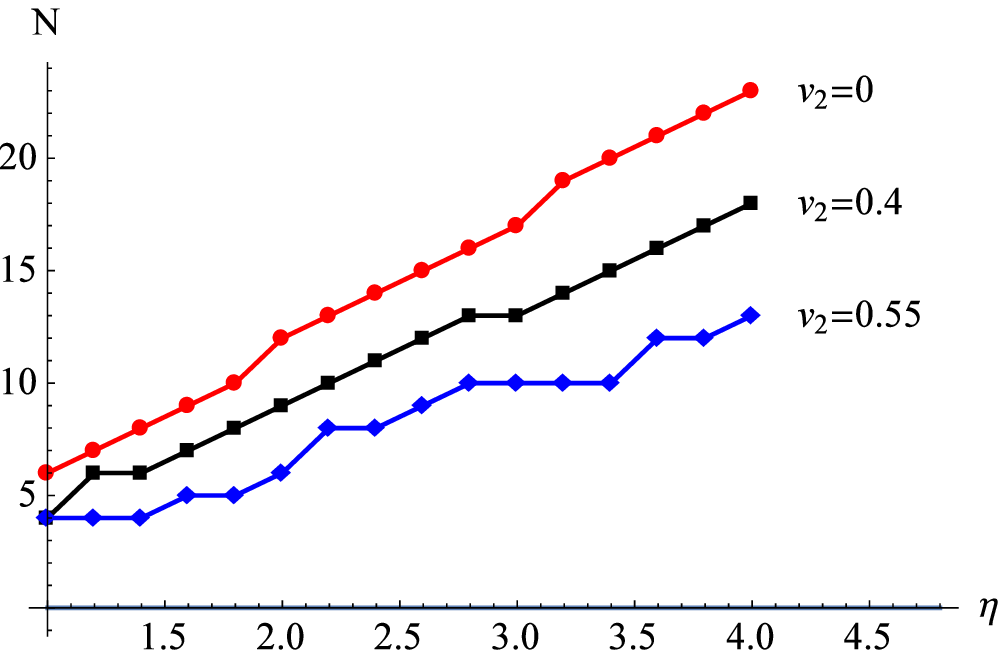}}\\
\caption {  \label{fig_PL_v2}
   The relationship of the number $N$ of the resonant KK modes with the Yukawa coupling $\eta$ for different values of $D$ and $v_2$.
   The parameters are set to $k=1, v_1=v_3=1$, $D=10,~20$, and $v_2=0,~0.4,~0.55$.}
\end{figure}
%%%%%%%%%%%%%%%%%%%%%%%%%%%%%%%%%%%%%%%%%%%%%%%%%%%%%%%%%%%%%%%

{\em a) The effect of $\eta$.}
 The curves for number of the resonances $N$ versus coupling coupling $\eta$ are shown in Fig.~\ref{fig_PL_v2}, where we have set the width of the brane array  $D$ to be $10$ in Fig.~\ref{fig_PL_v2a} and $20$ in Fig.~\ref{fig_PL_v2b}. It is clear that the number of the resonant KK modes increases with the Yukawa coupling constant $\eta$. This is because the height of potential well increases with the coupling constant $\eta$ as shown in Fig.~\ref{fig_potential_two_three_brane_b}. So there are more and more massive resonant modes by strengthening the Yukawa interaction, and this is a well-known result revealed in previous literatures.

{\em b) The effect of $v_i$.}
The figure~\ref{fig_PL_v2b} shows that the resonance number $N$ decreases with the amplitude parameter $v_2$. This is due to the fact that the height of potential barriers will decrease with $v_2$, as shown in Fig.~\ref{fig_potential_two_three_brane_c}.

{\em c) The effect of $D$.}
By comparing Fig.~\ref{fig_PL_v2a} with Fig.~\ref{fig_PL_v2b},  it can be concluded that the number of the resonant KK modes increases with the total width $D$ of the brane array. This can be understood from the fact that
the width of the potential well is proportional to the total width $D$ of the brane array, and the  broader potential well is, the more resonant KK modes are produced.

If the parameter $v_2$ vanishes, the middle sub-brane disappears, so the curves referring to a null $v_2$ correspond to a two sub-brane system. It indicates that the number of the sub-branes will affect the resonance number. So next we analyze this effect in detail. It is note that one can increase the numbers of the sub-banes with a fixed total width $D$ of the brane array, or with a fixed brane spacing $d$. However, since the width of the potential well is approximate to the width $D$ of the brane array, if one increases the brane array with a fixed brane spacing $d$, the width and depth of potential well will change both. Therefore, we investigate the effect of $n$ simply with a fixed total width $D$.

\begin{figure}[htbp]
\centering
\subfigure[]{\label{fig_VL_N_1a}
\includegraphics[width=6cm]{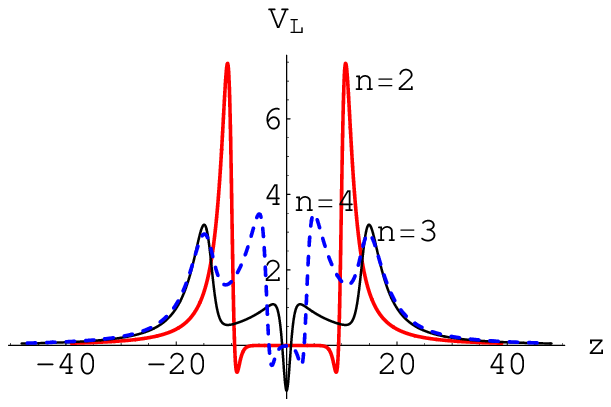}}
\subfigure[]{\label{fig_VL_N_1b}
\includegraphics[width=6cm]{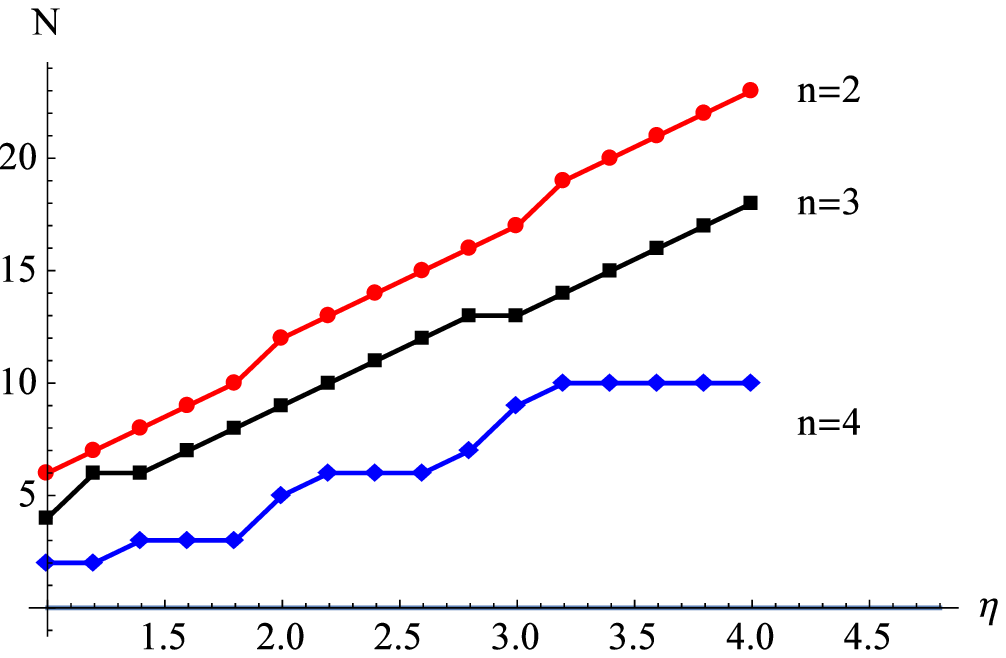}}
\caption{ \label{fig_VL_N}
 The effective potentials $V_{L}$ and the relationship of the number $N$ of the resonant KK modes with the Yukawa coupling $\eta$ for different values of $n$ and fixed $D$.
  The parameters are set to $k=1$ and $D=20$, $(v_1,v_2)=(1,1)$ for $n=2$,
  $(v_1,v_2,v_3)=(1,0.4,1)$ for $n=3$, and $(v_1,v_2,v_3,v_4)=(1,0.4,0.4,1)$ for $n=4$.}
\end{figure}

{\em d) The effect of $n$ with fixed total width $D$.}
We fix the total width $D=20$, increase the coupling constant $\eta$ from $1$ to $4$ and change the number $n$ of the sub-branes, the brane spacing reads $d=D/(n-1)$.
The effective potential well of the brane array and the relationship of the resonance number $N$ with the coupling constant $\eta$ for different values of $n$ and fixed $D$ is plotted in Fig.~\ref{fig_VL_N}.
Obviously, the resonance number increases with the coupling constant  $\eta$, no matter how much the number of sub-branes is, but decreases when the number $n$ of the sub-branes increases (see Fig.~\ref{fig_VL_N_1b}). This is because the depth of potential well decreases with the number $n$ as shown in Fig.~\ref{fig_VL_N_1a}.
%%%%%%%%%%%%%%%%%%%%%%%%%%%%%%%%%%%%%%%%%%%%%%%%%%%%%%%%%%%%%%
%%%%%%%%%%%%%%%%%%%%%%%%%%%%%%%%%%%%%%%%%%%%%%%%%%%%%%%%%%%%%%%

\subsection{Mass spectrum of resonant fermions}

In order to investigate the properties of the fermion resonant
mass spectra on the brane array, we have to input the number $n$ of the sub-branes. However, we found the properties are nothing new in models of one and two sub-branes, but there are some novel properties emerging from three or four sub-branes system.

%%%%%%%%%%%%%%%%%%%%%%%%%%%%%%%%%%%%%%%%%%%%%%%%%%%%%%%%%%%%%%
\begin{figure}[htbp]
\centering
\subfigure[]{\label{fig_potential_abnormal_resonance_numbera}
\includegraphics[width=6cm]{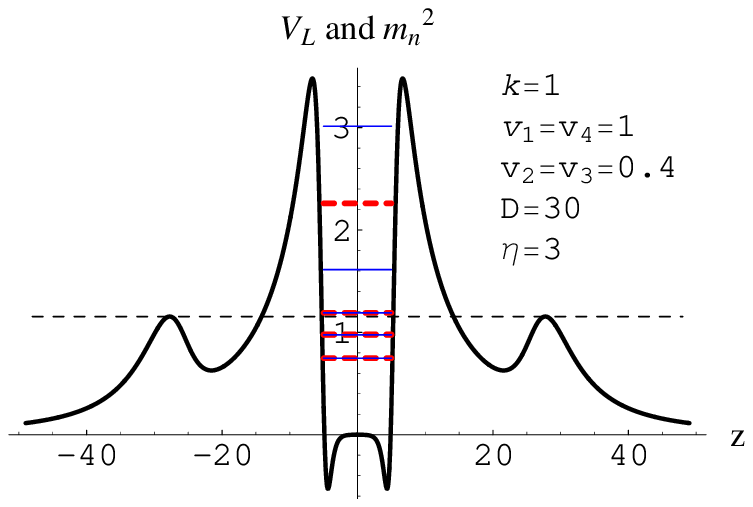}}
\subfigure[]{  \label{fig_potential_abnormal_resonance_numberb}
\includegraphics[width=6cm]{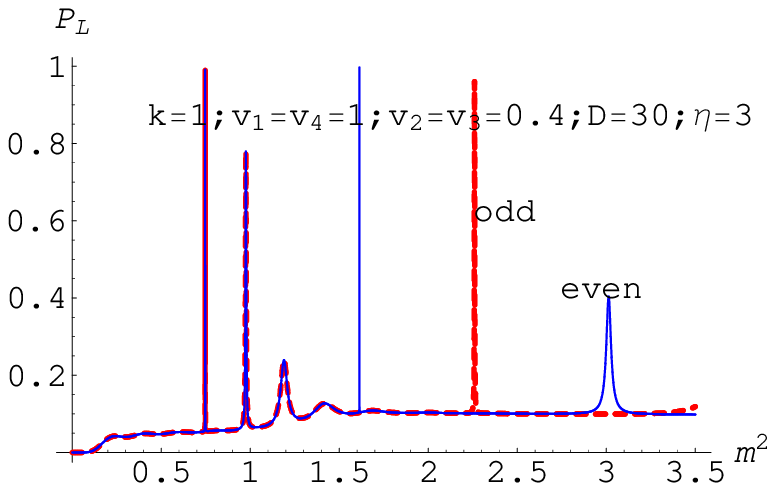}}\\
\subfigure[local enlarged drawing of (b)]{  \label{fig_potential_abnormal_resonance_numberc}
\includegraphics[width=6cm]{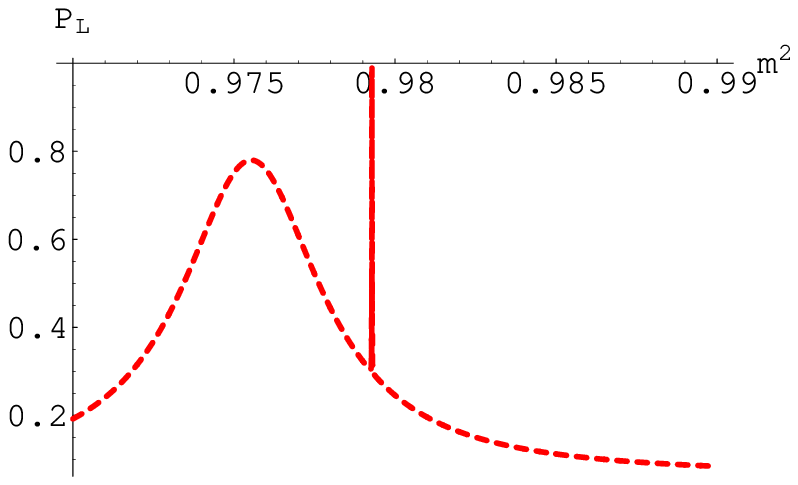}}
\subfigure[local enlarged drawing of (b)]{  \label{fig_potential_abnormal_resonance_numberd}
\includegraphics[width=6cm]{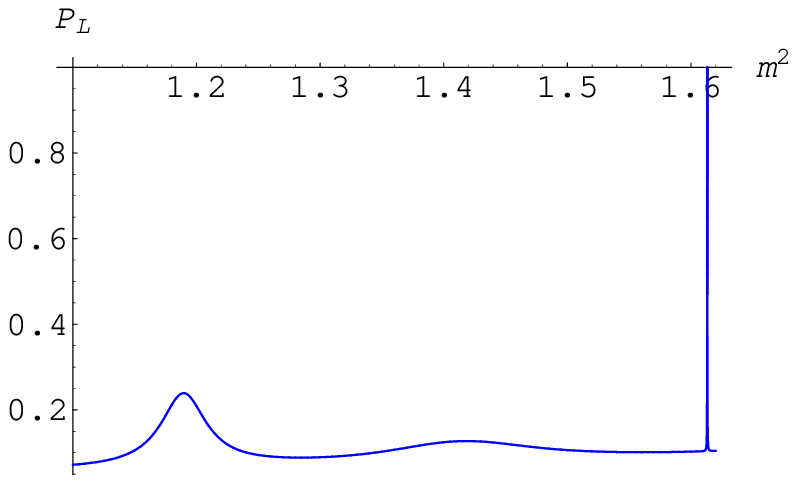}}
\caption{ \label{fig_potential_abnormal_resonance_numbera_4brane}
\small
The plots of effective potential $V_{\rm L}(z)$ (black lines), resonance spectrum $m_n^2$ (red dashed lines for parity-odd and blue thin lines for parity-even), and the probability $P_{\rm L}{(m^2)}$ of the left-chiral fermion KK modes $f_{{\rm L}n}(z)$ for the four sub-brane model.  }
\end{figure}
%%%%%%%%%%%%%%%%%%%%%%%%%%%%%%%%%%%%%%%%%%%%%%%%%%%%%%%%%%%%%%%

Firstly, we focus on a symmetric four sub-brane model, with the parameters are set to be $n= 4, D= 30, v_1 = v_4 = 1,  v_2 = v_3 = 0.4$ and $\eta=3$. The corresponding effective potential and relative probability $P_L$
are shown in  Fig.~\ref{fig_potential_abnormal_resonance_numbera_4brane}. 
Any peak in the curves of relative probability refers to a resonant state. The mass spectrum of the resonant states read from  Fig.~\ref{fig_potential_abnormal_resonance_numberb} is listed in Tab. \ref{tab1}.  It is interesting that there are three resonant states with the masses \{0.7472, 0.9755, 1.1897\} in both parities. Therefore, the resonant states are doubly-degenerate for the three masses. To the best of our knowledge, this does not happen for the cases of one and two sub-branes, where all the resonant states are non-degenerate and possess either odd-parity or even-parity.

\begin{table}
\begin{center}
\begin{tabular}{|c|c|c|c|c|c|c|}
\hline
$n$      &parity  &$m^2$     &$m$       &$\Gamma$     &$\tau$\\
\hline
$1$      &both    &$0.7472$  &$0.8644$  &$0.000294$   &$3403$  \\
\hline
$2$      &both    &$0.9755$  &$0.9877$  &$0.002891$   &$346$  \\
\hline
$3$      &odd    &$0.9793$  &$0.9896$  &$0.0000045$   &$222379$  \\
\hline
$4$      &both    &$1.1897$  &$1.0907$  &$0.031021$   &$32$  \\
\hline
$5$      &even    &$1.6132$  &$1.2701$  &$0.000077$   &$12960$  \\
\hline
$6$      &odd    &$2.2598$  &$1.5033$  &$0.000948$   &$1055$ \\
\hline
\end{tabular}
\end{center}
\caption{The parity, mass $m$,  width $\Gamma$, and lifetime $\tau$ of resonances of the left-handed fermions for the four sub-brane model, where ``both" represents the doubly-degenerate states with an odd-parity fermion and an even-parity fermion. } \label{tab1}
\end{table}

It is well-known that all the bound states of one-dimensional Schr\"odinger equation with an even potential must be either odd or even parity. In our case, there is only one non-degenerate massless bound state with even-parity for either left-chiral or right-chiral fermions. For the unbounded states, due to the potential asymptotically vanishing far to the left and far to the right, there are two solutions for each eigenvalue $m^2$, and the two solutions can be written as one odd and one even functions under the reflection of extra dimension coordinate. So the mass spectrum of massive fermions is continuous and doubly-degenerate. However, by imposing either kind of boundary conditions in Eq.~(\ref{Boundary_conditions}), one picks up either odd or even eigenfunctions. Then the condition of being resonance picks up some eigenvalues and leads the resonant mass spectrum being discrete. Generally, if one eigenfunction with a certain parity (either odd or even) possess a low transmission coefficient through a potential barriers with certain mass (or sharply peaked at the curves of relative probability $P_L$), the other degenerate eigenfunction with the opposite parity can not possess a low transmission simultaneously. So generally, the resonant states are non-degenerate. This is just the case in one and two sub-brane systems.

    \begin{figure*}[!htb]
    \subfigure[$m^2=0.7472$]{\label{fig_potential_abnormal_resonant_wave_function_1}
    \includegraphics[width=5cm]{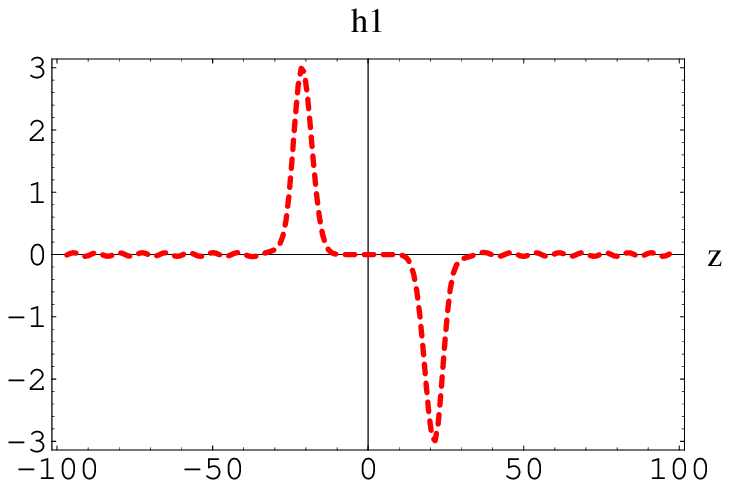}}
    \subfigure[$m^2=0.7472$]{\label{fig_potential_abnormal_resonant_wave_function_2}
    \includegraphics[width=5cm]{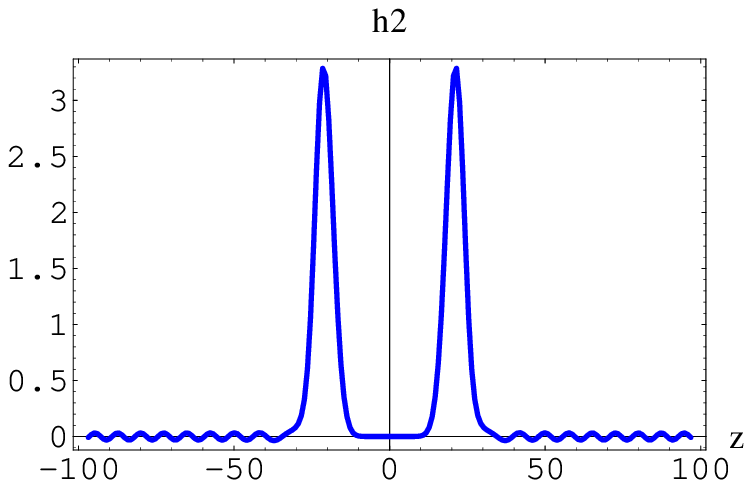}}
    \subfigure[$m^2=0.9755$]{\label{fig_potential_abnormal_resonant_wave_function_3}
    \includegraphics[width=5cm]{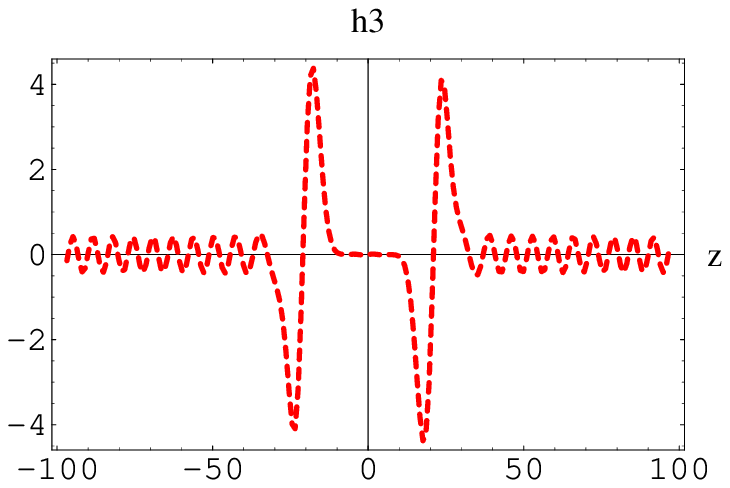}}
    \subfigure[$m^2=0.9755$]{\label{fig_potential_abnormal_resonant_wave_function_4}\\
    \includegraphics[width=5cm]{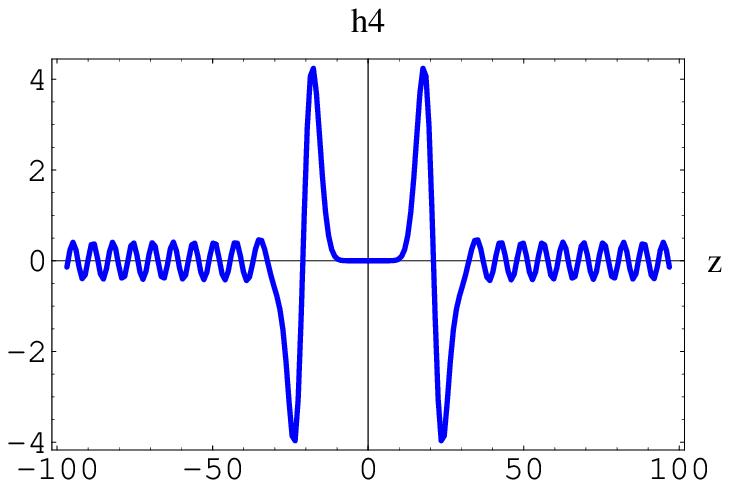}}
    \subfigure[$m^2=0.9793$]{\label{fig_potential_abnormal_resonant_wave_function_5}
    \includegraphics[width=5cm]{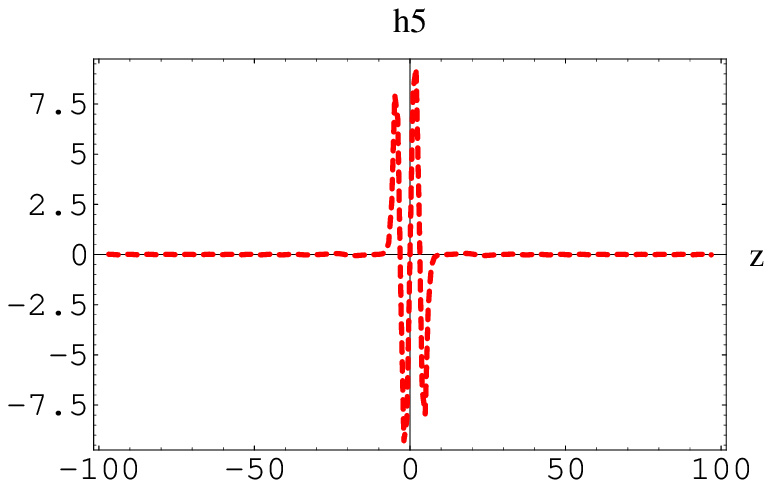}}
    \subfigure[$m^2=1.1897$]{\label{fig_potential_abnormal_resonant_wave_function_6}
    \includegraphics[width=5cm]{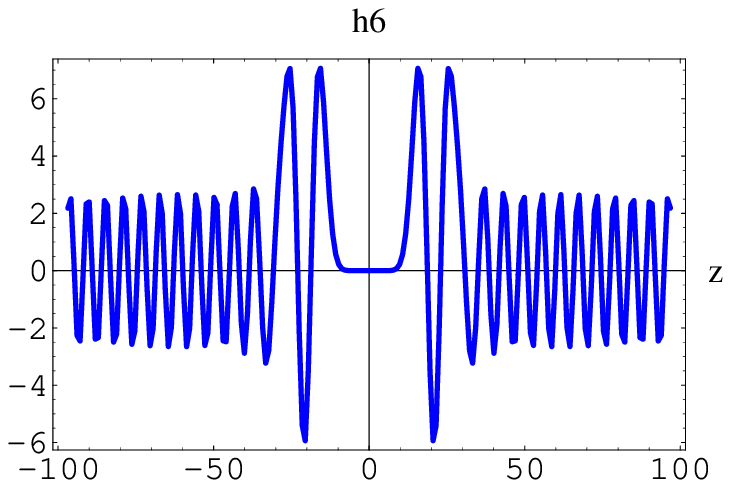}}
    \subfigure[$m^2=1.1897$]{\label{fig_potential_abnormal_resonant_wave_function_7}
    \includegraphics[width=5cm]{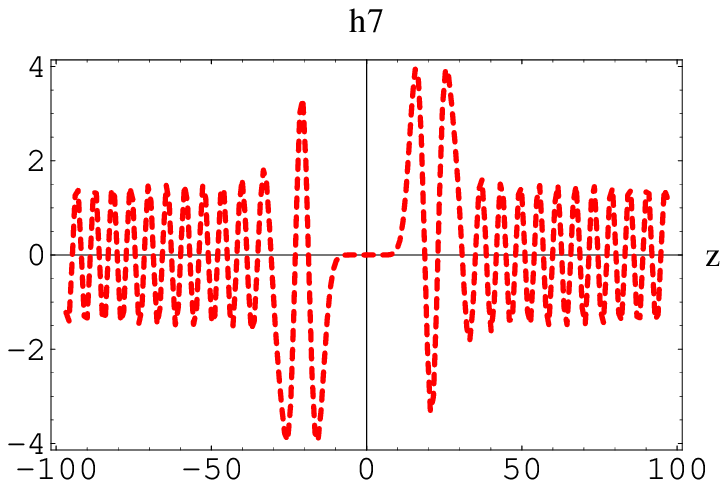}}
    \subfigure[$m^2=1.6132$]{\label{fig_potential_abnormal_resonant_wave_function_8}
    \includegraphics[width=5cm]{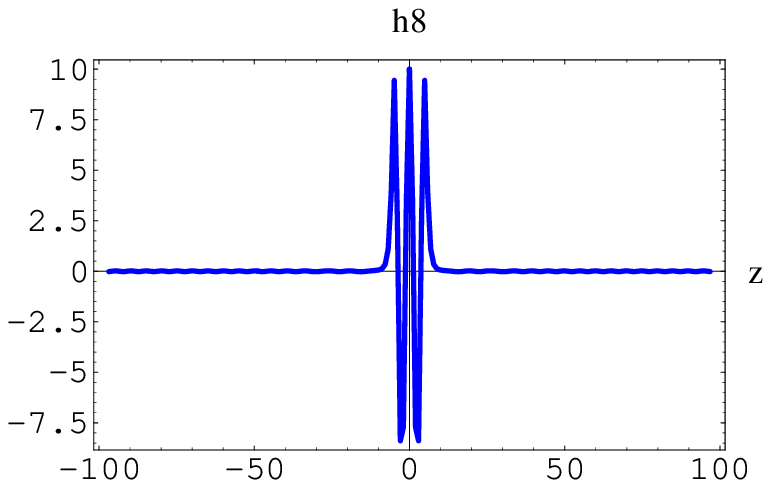}}
    \vskip -0mm \caption{The fermion resonant KK modes $h_1(z)-h_8(z)$ of the four sub-brane model with the parameters set to $k= 1,~v_1=v_4= 1,~v_2=v_3=0.4,~D=30,$ and $~\eta=3$. The odd and even resonance  KK modes are denoted by the red dashed lines and the blue continuous lines, respectively.}
    \label{fig_potential_abnormal_resonant_wave_function}
    \end{figure*}

However, for the case of four sub-brane system, there are four peaks in the effective potential as is shown in Fig.~\ref{fig_potential_abnormal_resonance_numbera}. Comparing with the effective potential of two sub-brane system shown in Fig.~\ref{fig_potential_two_three_brane_c} with $v_2=0$, it is clear that there are two additional shallower potential wells on the two shoulders of potential barriers, where bottom of the wells is $V_L=0.62746$ and top is $V_L=1.1541$. The Fig.~\ref{fig_potential_abnormal_resonance_numbera} shows that the eigenvalues \{0.7472, 0.9755, 1.1897\} of the three resonant states are almost located in the range (0.62746, 1.1541) of the shoulder potential wells. Therefore, it implies that the three doubly-degenerate states may be closely related to the two shoulder potential wells. And this can be verified from the wave functions presented in Fig.~\ref{fig_potential_abnormal_resonant_wave_function}, where the odd and even wave functions for eigenvalue $m^2=0.7472$ are plotted in Figs.~\ref{fig_potential_abnormal_resonant_wave_function_1} and \ref{fig_potential_abnormal_resonant_wave_function_2} respectively, the odd and even wave functions for $m^2=0.9755$ in Figs.~\ref{fig_potential_abnormal_resonant_wave_function_3} and \ref{fig_potential_abnormal_resonant_wave_function_4} respectively, and the old and even wave function for $m^2=1.1897$ in Figs.~\ref{fig_potential_abnormal_resonant_wave_function_6} and \ref{fig_potential_abnormal_resonant_wave_function_7} respectively. By ascertaining the coordinate ranges of those peaks, it is clear that all the resonant states are quasi-localized in the shoulder potential wells. Specifically,  as shown in Figs.~\ref{fig_potential_abnormal_resonant_wave_function_1} and \ref{fig_potential_abnormal_resonant_wave_function_2}, since the peak of wave functions has no nodes locally in the shoulder potential wells, they are the first resonant state of the shoulder wells. Similarly,  Figs.~\ref{fig_potential_abnormal_resonant_wave_function_3} and \ref{fig_potential_abnormal_resonant_wave_function_4} indicate that the modes is the second resonant state with one node locally in the shoulder wells, and Figs.~\ref{fig_potential_abnormal_resonant_wave_function_6} and \ref{fig_potential_abnormal_resonant_wave_function_7} indicate the modes are the third resonant state with two nodes locally in the shoulder wells. Finally, an even solution is equipped by a symmetric combination of the resonant states of two shoulder wells, and an odd solution by an antisymmetric combination of the resonant states.

However, as illustrated in Figs.~\ref{fig_potential_abnormal_resonant_wave_function_5} and \ref{fig_potential_abnormal_resonant_wave_function_8}, the non-degenerate modes are manifest quasi-localized in the center potential well. Here, it is note that since the modes in Figs.~\ref{fig_potential_abnormal_resonant_wave_function_5} and \ref{fig_potential_abnormal_resonant_wave_function_8} have three and four nodes, respectively. Therefore, there is an even non-degenerate resonance with two nodes and an odd non-degenerate resonance with one node missing in  Fig.~\ref{fig_potential_abnormal_resonance_numberb}. This is due to their resonant widths are too narrow to be picked out by the precision we used. Further, since the center well is much deeper than the two shoulder wells, the lifetimes of resonances in the center well are much longer than that in the shoulder wells. This is exactly the case presented in table~\ref{tab1}, where the lifetimes of non-degenerate resonances are much longer than that of their adjacent degenerate resonances, and it can also be seen clearly from Figs.~\ref{fig_potential_abnormal_resonance_numberc} and \ref{fig_potential_abnormal_resonance_numberd}, where the non-degenerate resonant modes with eigenvalues $m^2=0.9793$ and $m^2=1.6132$ appear as much sharper peaks than the adjacent degenerate resonances in the curves of relative probability.

%%%%%%%%%%%%%%%%%%%%%%%%%%%%%%%%%%%%%%%%%%%%%%%%%%%%%%%%%%%%%%
\begin{figure}[htbp]
\centering
\subfigure[]{\label{fig_abnomal_resonance_3branea}
\includegraphics[width=7cm]{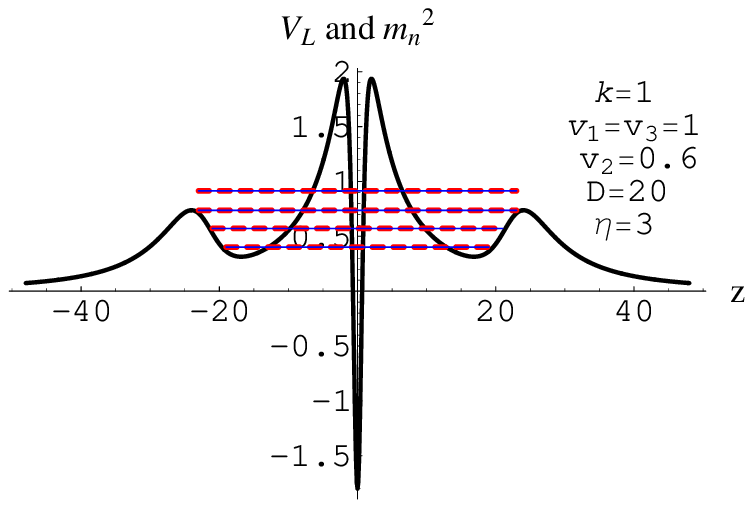}}
\subfigure[]{  \label{fig_abnomal_resonance_3braneb}
\includegraphics[width=7cm]{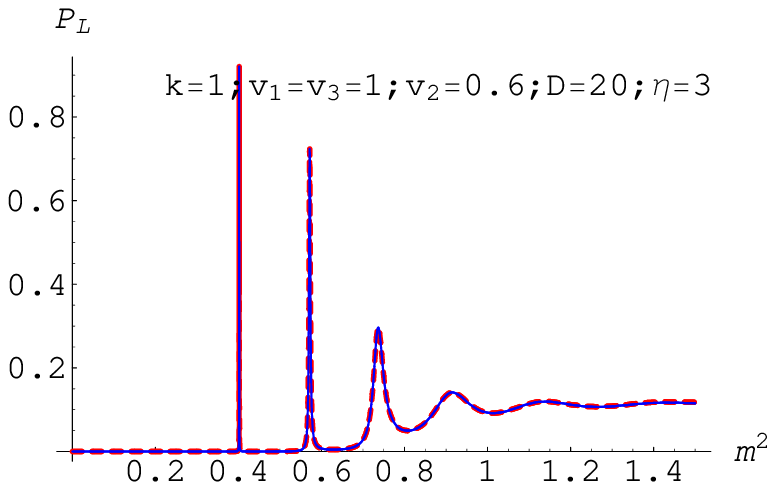}}\\
\caption{ \label{fig_abnomal_resonance_3brane}
\small The plots of effective potential $V_{\rm L}(z)$ (black lines), resonance spectrum $m_n^2$ (red dashed lines for parity-odd and blue thin lines for parity-even), and the probability $P_{\rm L}{(m^2)}$ of the left-chiral fermion KK modes $f_{{\rm L}n}(z)$ for the three sub-brane model.  }
\end{figure}
%%%%%%%%%%%%%%%%%%%%%%%%%%%%%%%%%%%%%%%%%%%%%%%%%%%%%%%%%%%%%%%

\begin{figure*}[!htb]
    \subfigure[$m^2=0.4019$]{\label{fig_three_brane_wave_function_1}
    \includegraphics[width=5cm]{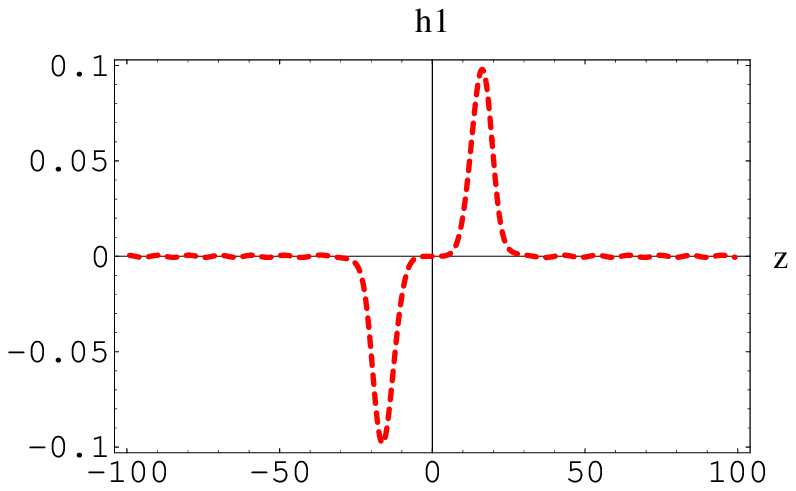}}
    \subfigure[$m^2=0.4019$]{\label{fig_three_brane_wave_function_2}
    \includegraphics[width=5cm]{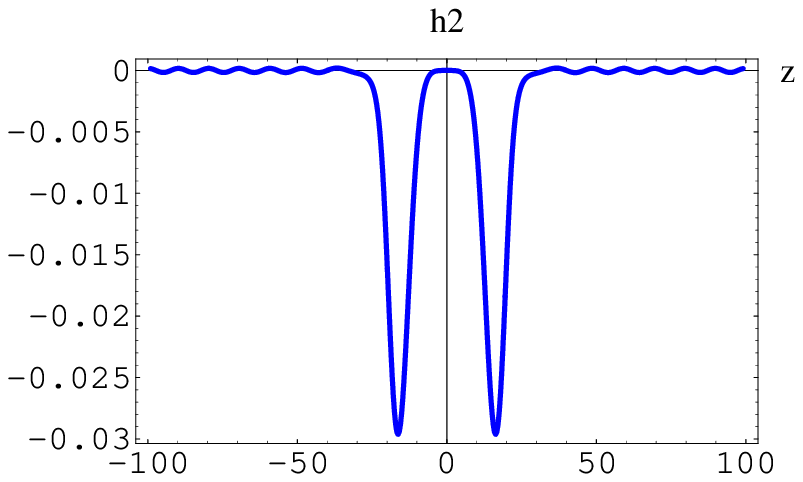}}
    \subfigure[$m^2=0.5716$]{\label{fig_three_brane_wave_function_3}
    \includegraphics[width=5cm]{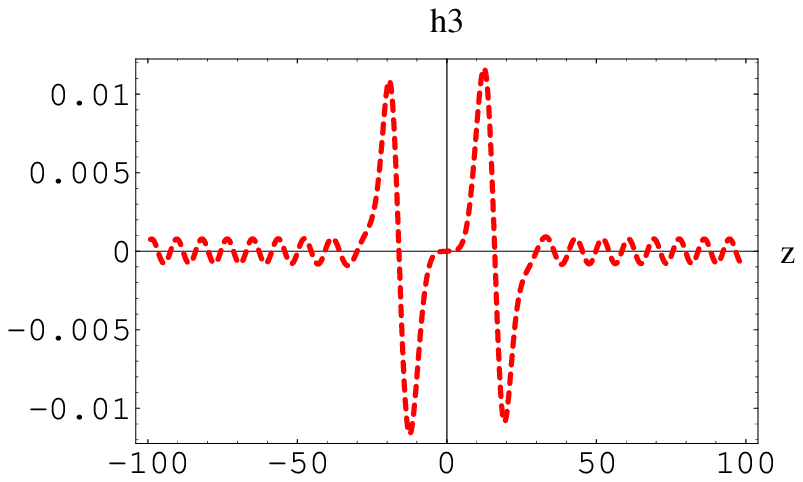}}
    \subfigure[$m^2=0.5716$]{\label{fig_three_brane_wave_function_4}\\
    \includegraphics[width=5cm]{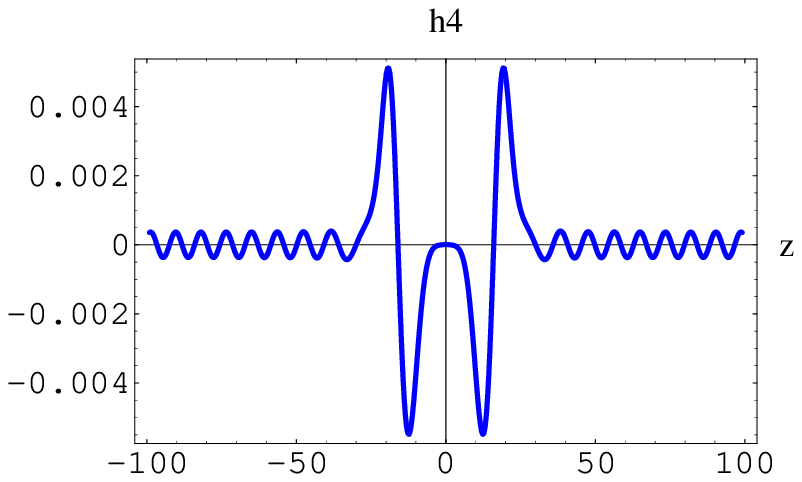}}
    \subfigure[$m^2=0.7367$]{\label{fig_three_brane_wave_function_5}
    \includegraphics[width=5cm]{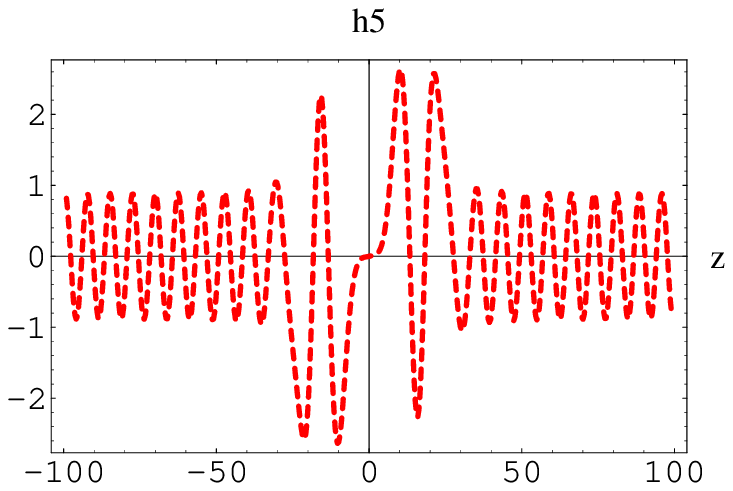}}
    \subfigure[$m^2=0.7367$]{\label{fig_three_brane_wave_function_6}
    \includegraphics[width=5cm]{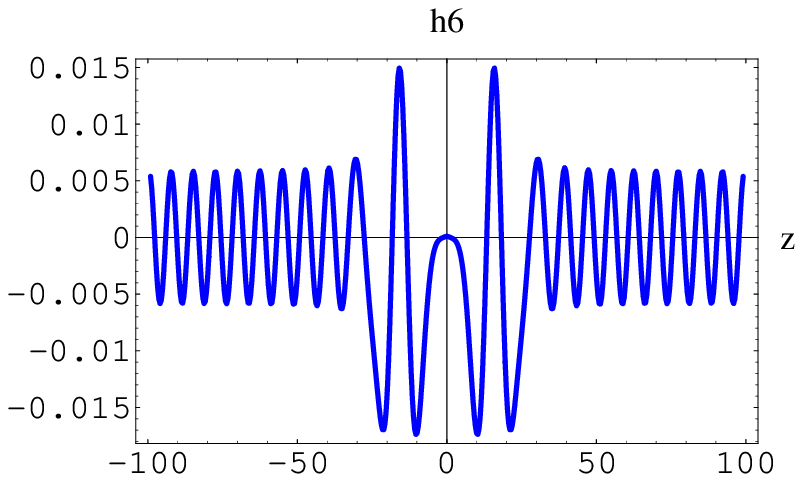}}
    \subfigure[$m^2=0.9147$]{\label{fig_three_brane_wave_function_7}
    \includegraphics[width=5cm]{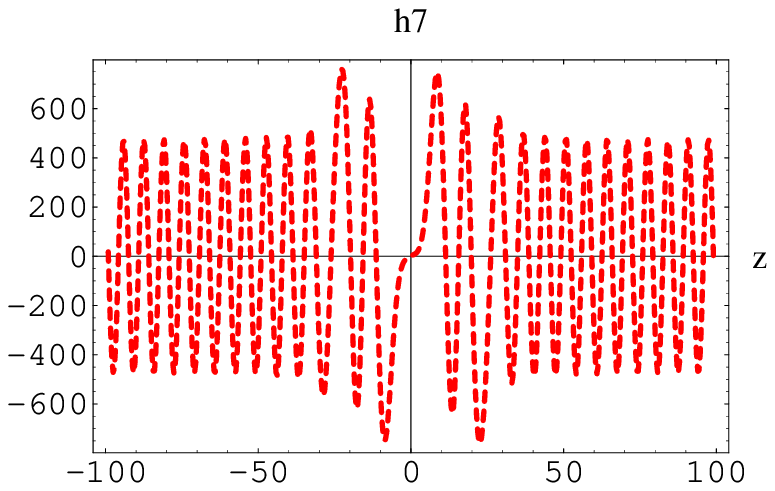}}
    \subfigure[$m^2=0.9147$]{\label{fig_three_brane_wave_function_8}
    \includegraphics[width=5cm]{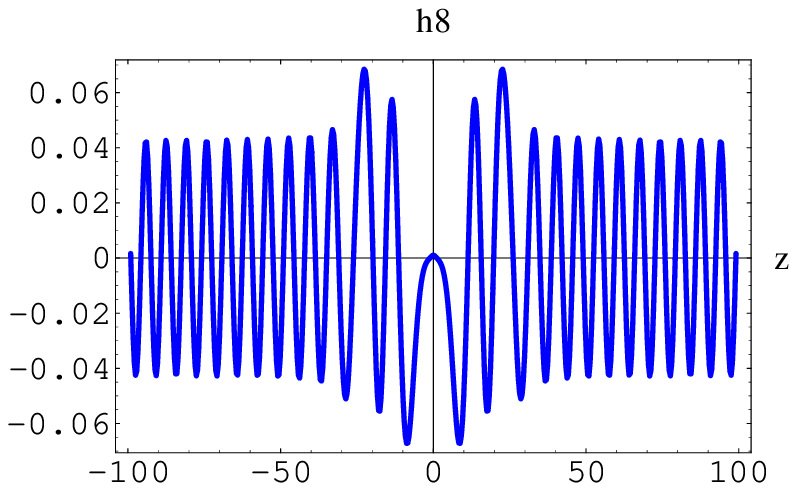}}
    \vskip -0mm \caption{The fermion resonant KK modes $h_1(z)-h_8(z)$ of the three sub-brane model with the parameters set to $k= 1,~v_1=v_3= 1,~v_2=0.6,~D=20,$ and $~\eta=3$. The odd and even resonant KK modes are denoted by the red dashed lines and the blue continuous lines, respectively.}
    \label{fig_three_brane_resonant_wave_function}
    \end{figure*}

 For the case of three sub-brane system, if there appears two shoulder potential wells, such as the case shown in Fig.~\ref{fig_abnomal_resonance_3branea}, it may also exist doubly-degenerate quasi-localized resonant states (see Fig.~\ref{fig_abnomal_resonance_3braneb}). The corresponding resonant mass spectrum, widths and life times are shown in Tab.~\ref{tab2}. As shown in Fig.~\ref{fig_abnomal_resonance_3brane}, all the four resonant states are degenerate. On the other hand, it is clear that all the resonances are quasi-localized in the shoulder potential wells by observing the wave functions illustrated in Fig.~\ref{fig_three_brane_resonant_wave_function}. This is possibly because the central potential well is too narrow to support a  resonance. So in this case, the lifetime of the resonances declines monotonously as resonant mass increases. This is different from the case of four sub-brane system, where there are both states localized in central well and shoulder wells. 

\begin{table}
\begin{center}
\begin{tabular}{|c|c|c|c|c|c|c|}
\hline
$n$      &parity  &$m^2$     &$m$       &$\Gamma$     &$\tau$\\
\hline
$1$      &both    &$0.4019$  &$0.6339$  &$0.00159$   &$630$  \\
\hline
$2$      &both    &$0.5716$  &$0.7560$  &$0.00319$   &$314$  \\
\hline
$3$      &both    &$0.7367$  &$0.8584$  &$0.0317$   &$32$  \\
\hline
$4$      &both    &$0.9147$  &$0.9565$  &$0.0774$   &$13$  \\
\hline
\end{tabular}
\end{center}
\caption{The parity, mass $m$,  width $\Gamma$, and lifetime $\tau$ of resonances of the left-handed fermions for the three sub-brane model, where ``both" represents the doubly-degenerate states with an odd-parity fermion and an even-parity fermion.} \label{tab2}
\end{table}

\begin{figure}[htbp]
\centering
\subfigure[five sub-brane model]{
\includegraphics[width=7cm]{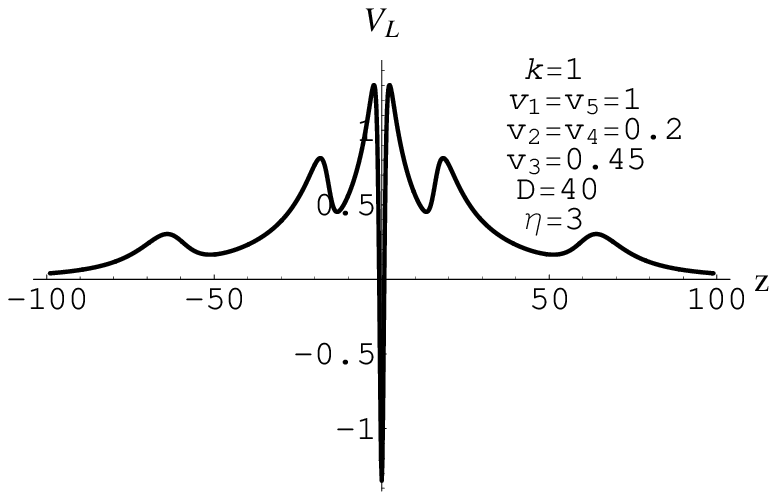}}
\subfigure[six sub-brane model]{ 
\includegraphics[width=7cm]{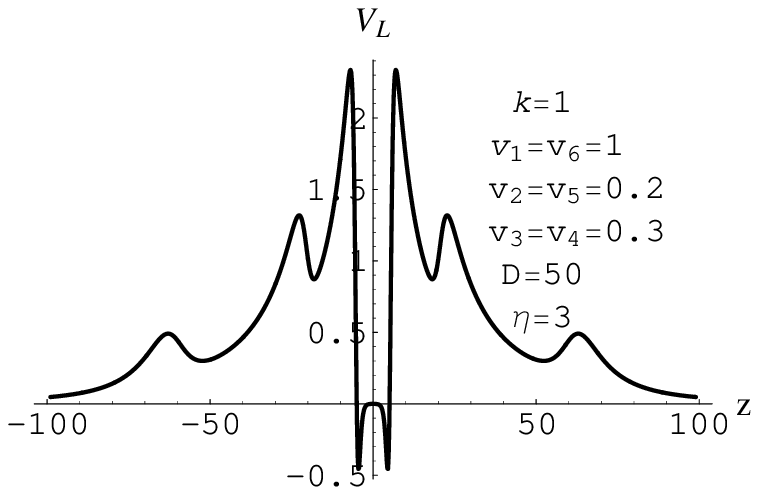}}\\
\caption{ \label{fig_abnomal_resonance_56branes}
\small The plots of effective potentials $V_{\rm L}(z)$ for the five and six sub-brane models.  }
\end{figure}

These new properties emerge in the cases of three and four sub-branes, and there is no doubt that it will be more complicated in the cases of more sub-branes. Such as the cases of five and six sub-branes illustrated in Fig.~\ref{fig_abnomal_resonance_56branes},  there could be more shoulder potential wells, so there are more series of degenerate resonant modes. Especially, by comparing their potential shapes, we can conclude that the properties of even sub-brane systems are more similar to that of four sub-branes, and the properties of odd sub-brane systems are more similar to that of three sub-branes.  

Further, for the usual single-wall braneworld model (simply fixing $n=1$), the mass splitting of resonant fermions is proportional to the parameter $v_1$ with mass-dimension one, whose inverse $1/v_1$ is proportional to brane thickness. Because there is no more mechanism to provide a new mass scale, the parameter is usually set to be the five-dimensional fundamental scale, which is equal to the four-dimensional Planck mass. So it is hopeless to probe the signal of these resonant fermions. However, by including multi sub-branes in the brane array, the resonant fermions are quasi-localized between the whole sub-branes, the mass splitting is related to inverse of the total width of brane array $D=|y_n-y_1|$, i.e., proportional to the new mass scale $M_D\equiv1/D$. By including more and more sub-branes,  $M_D$ would reduce to a lower and lower mass scale, such as a few TeV. So this novel phenomenon emerging in brane array could be potentially interesting in phenomenology.
%%%%%%%%%%%%%%%%%%%%%%%%%%%%%%%%%%%%%%%%%%%%%%%%%%%%%%%%%%%%%%
%%%%%%%%%%%%%%%%%%%%%%%%%%%%%%%%%%%%%%%%%%%%%%%%%%%%%%%%%%%%%%%

\section{Conclusions}
\label{sec:conclusion}

In this work, we constructed the multi-wall braneworld model with an arbitrary number of bulk scalar fields, and gave two analytic solutions by imposing the polynomial superpotential and modified sine-Gordon superpotential, respectively. Then, for the massless fermion, we studied its localization property on the brane array and came to the conclusion that the massless fermion is localized between the two outermost sub-branes. For the massive resonant fermions, we investigated the effects of some parameters on the number $N$ of resonances.  We found that the number of resonances increases with the coupling constant $\eta$ and the total width $D$ of the brane array, but decreases with the amplitude parameter $v_2$ of the middle sub-brane. Moreover, if we fix the total width $D$, the number of resonances decreases with the number of the sub-branes, while if we fix the brane spacing $d$, the dependency between the number of resonances and the number of the sub-branes is not simply monotonous. Finally, we discussed the mass spectrum of resonant fermions. It is significant that there are doubly-degenerate resonant modes for the cases of
three and four sub-brane systems with some sets of parameter values. However, this does not happen for the cases of one and two sub-branes, where all the resonant states are non-degenerate. Besides, the non-degenerate resonant modes appear as much sharper peaks than the adjacent degenerate resonant modes in the curves of relative probability. By setting the mass scale $M_D$ not to be far above TeV, the novel phenomenon emerging in brane array could be potentially interesting in phenomenology, which is left for our future consideration.

\ack
We would like to thank Prof. Yu-Xiao Liu for helpful discussion. This work was supported by the National Natural Science Foundation of China (Grants No. 11705070 and No. 11747021). K. Yang acknowledges the support of ``Fundamental Research Funds for the Central Universities" under Grant No. XDJK2019C051.

\end{document}